\documentclass[11pt,tightenlines,eqsecnum,floats,aps,nofootinbib,prd,showpacs]{revtex4}
\usepackage{hyperref}
\usepackage{amsmath,amssymb,amsfonts,amsthm,amscd}
\usepackage{graphicx}
\usepackage{enumerate}
\usepackage{colordvi}
\usepackage{units}
\usepackage{epsfig}
\usepackage{natbib}
\usepackage{enumerate}
\usepackage{colordvi}
\usepackage{multirow}
\usepackage{afterpage}
\usepackage{pdflscape}
\usepackage[caption=false]{subfig}
\usepackage{braket}

\def\be{\begin{equation}}
\def\ee{\end{equation}}
\def\ba{\begin{eqnarray}}
\def\ea{\end{eqnarray}}

\def\L{\mathcal{L}}
\def\H{\mathcal{H}}
\def\p{P_\phi}
\def\f{\frac}

\DeclareMathOperator{\sech}{sech}
\DeclareMathOperator{\csch}{csch}
\DeclareMathOperator{\arctanh}{arctanh}

\DeclareMathOperator{\arccosh}{arccosh}

\begin{document}

\title{T-Model Inflation and Bouncing Cosmology}

\author{David Sloan}
\email{d.sloan@lancaster.ac.uk}
\affiliation{Consortium for Fundamental Physics, Physics Department, Lancaster University, Lancaster LA1 4YB, UK}

\author{Konstantinos Dimopoulos}
\email{konst.dimopoulos@lancaster.ac.uk}
\affiliation{Consortium for Fundamental Physics, Physics Department, Lancaster University, Lancaster LA1 4YB, UK}

\author{Sotirios Karamitsos}
\email{sotirios.karamitsos@manchester.ac.uk}
\affiliation{Consortium for Fundamental Physics, Physics Department, Manchester University, Manchester M13 9PL, UK}

\begin{abstract}
 
We examine the dynamics of a closed cosmology whose matter source is that of a conformally coupled scalar field with a broken ${\rm SO}(1,1)$ symmetry, which correspond to the {$\alpha$-attractors} proposed by Linde and Kallosh. Following a field redefinition, such models give rise to ``T-model'' inflationary potentials, whose dynamics provide both an inflationary phase and a classical bounce. We show that the universe can undergo bounces far from the regime of quantum gravity (i.e. at energy densities lower than the Planck density). We analyse perturbations on this background with particular attention given to the effects of a double-bouncing scenario (with rapid recollapse between bounces) on the long wavelength modes. We demonstrate that the predictions of such models agree well with observations and might explain the suppression of power in the low multiples of the CMB.
\end{abstract}

\maketitle

\section{Introduction}
 
The inflationary paradigm has become a central feature of modern cosmology since it both addresses conceptual issues such as the flatness, horizon and entropy problems and also provides a natural mechanism by which primordial fluctuations can seed structure formation
\cite{Starobinsky:1980te,Sato:1981ds,Kazanas:1980tx,Guth:1980zm} (for a review see Ref.~\cite{Baumann:2009ds}). This is achieved by the \emph{nearly} de Sitter geometry of the expanding early universe, which generates an almost scale-invariant spectrum of perturbations. In order to achieve this, the dominant form of matter must have an equation of state $p \approx -\rho$, i.e. $w=-1$, that of a cosmological constant. In order for inflation to end, $w$ must increase in order to allow other forms of matter to contribute significantly to the evolution of the universe. Typically, models invoke scalar fields to realize such an evolution; when the field is potential-dominated, it behaves as a cosmological constant. As it descends down the potential and becomes kinetically dominated however, it acts more like a stiff fluid ($w=1$). In a closed cosmology, the $k>0$ Friedmann-Lem\^aitre-Robertson-Walker (FLRW) model wherein the spatial manifold is a 3-sphere, the same accelerated expansion can lead to a bounce from a collapsing universe to an expanding one. It is therefore natural to ask whether the same mechanism can give rise to both phenomena. Qualitatively, the collapse and the expansion impose requirements that are ``mirror images'' of one another on the field. In a contracting branch, the field must gain energy faster than other energy sources, yet become almost constant under evolution at small length scales. In the expanding branch, the opposite is true; the energy density should dissipate slowly only at small scales and faster as the universe expands. Typically, a scalar field with a suitable potential is invoked to play the role of the inflaton (the field responsible for inflation), as the dynamics of such a system allow for this dual role.

Observational data from the Planck satellite has shown both excellent agreement with the FLRW cosmological paradigm and place strict bounds on cosmological parameters \cite{Akrami:2018odb,Aghanim:2018eyx}. Taken together with the baryon acoustic oscillations, the size of the curvature constant, $\Omega_k$ is quite tightly constrained around zero -- the best fit parameter value gives $\Omega_k = -0.005^{+0.016}_{-0.017}$ under the parametrisation $\sum \Omega_i =1$. These bounds are likely to be improved by future observations. However, there is a limit which will be reached if the observations $|\Omega_k|<10^{-5}$ become degenerate with the effects of a curvature perturbation whose wavelength is greater than the Hubble horizon. Therefore, in the absence of a positive measurement of $\Omega_k$, the closed universe is impossible to rule out as a physical system. On the inflationary side, recent progress has stringently constrained models. The ratio of amplitudes of tensor perturbations to those of scalar perturbations, denoted $r$, together with the scalar spectral tilt $n_s$, strongly disfavour most monomial potentials, particularly the minimal quadratic and quartic models. Concave models which exhibit plateaus, however, appear to be in good correspondence with the observational measurements, with a large class giving rise to $n_s \approx 0.968$ and small $r$ (dependent on specific field configurations) \cite{Kallosh:2013yoa,Galante:2014ifa,Kallosh:2015lwa}. 

Since inflation is invoked to solve issues of fine-tuning of the initial parameters, it behoves us to use inflationary models which are in some sense ``natural'' or ``generic'' -- a solution to the horizon and flatness problems that requires a very specific choice of potential simply passes on the burden of explaining away fine-tuning \cite{Ijjas:2014nta,Brandenberger:2016uzh}. As the simplest of models are observationally excluded, we will consider inflationary models that arise naturally in the context of extended theories of gravity, and in particular from the conformal coupling of a scalar field to the gravitational action. In the case of the scalar field with a broken ${\rm SO}(1,1)$ symmetry, such theories give rise to ``T-models'' of inflation~\cite{Carrasco:2015rva} which (predominantly) asymptotically become plateaus for large field values. Although the quantitative results we present will be closely tied to this choice, the qualitative behaviour is not; the vast majority of inflationary systems (i.e. those which allow for accelerated expansion) also allow for a bounce in a closed universe. In passing, we mention that the $\alpha$-attractors proposal may also unify inflation with dark energy \cite{Peebles:1998qn,Carrasco:2015pla, Dimopoulos:2017zvq},
but we will not pursue this direction further in this paper.

The model we discuss differs from other bouncing models which introduce exotic matter, higher derivatives or quantum effects to overcome the gravitaional collapse. Rather than considering a universe which must begin at infinite extent, within this model the universe could begin from an initial singularity in the pre-bounce phase. Such a universe would expand to finite extent before recollapsing into a bounce at an energy density smaller than the Planck density. The bounce would be then followed by a long period of inflation which gives rise to the universe we observe. This would avoid the problems of having runaway growth of inhomogeneities which can spoil the bounce. This provides an entirely different scenario from the conventional collapsing universe of infinite size, and is also explored in reference \cite{Matsui:2019ygj}.

This paper is laid out as follows: in Section~\ref{conf} we describe the particular realization of the gravitational action that gives rise to this scenario, and establish how the symmetry gives rise to the plateaus of our inflationary potential. This is followed by Sections~\ref{background} and \ref{Phasespace}, in which we establish the necessary conditions for the universe to undergo a bounce and describe the space of bouncing solutions. In Section~\ref{analytic}, we exhibit a simple model that features an analytic bouncing solution.  Numerical simulations of the full dynamics are given in Section~\ref{numerics}. In Section~\ref{perturbation}, we discuss the role of growing perturbations in a homogeneous and isotropic background as the universe collapses. We discuss observational implications in Section~\ref{observ}. Finally, we give some concluding remarks in Section~\ref{conclusions}.

Unless otherwise noted, we will be working in reduced Planck units throughout this paper, in which $M_P^{-2} = 8\pi G= 1$ and $c=1$.

\section{Conformal Action}
\label{conf}
We begin by consider theories of conformally invariant gravity induced by the evolution of a dynamical gravitational constant, as first proposed by Brans and Dicke \cite{Brans:1961sx}. To give rise to a more interesting inflationary scenario, we will consider a class of matter couplings featuring two fields which initially exhibit a $\rm{SO}(1,1)$ symmetry. Such models are of pressing interest, as it has recently been found that the apparent fifth force they would exhibit is suppressed due to Weyl invariance, and therefore such models are cosmologically viable \cite{Ferreira:2016kxi}. Closely following the seminal analysis of~\cite{Kallosh:2013yoa}, let us consider a conformally invariant Lagrangian for gravity (non-minimally) coupled to two scalar fields, $\chi$ and $\psi$, given by:
\be \L = \sqrt{-g} \left[ \frac12
\partial_\mu \chi \partial^\mu \chi - \frac12
\partial_\mu \psi \partial^\mu \psi + \frac{1}{12}(\chi^2-\psi^2) R - \frac{1}{36} F\left(\frac{\psi}{\chi}\right) (\psi^2-\chi^2)^2 \right]. \label{conf1} \ee
In the above, $F$ is an arbitrary function which breaks the $SO(1,1)$ symmetry globally (if $F$ were constant instead, the action would exhibit a global de Sitter symmetry). Although the sign of the~$\psi^2$ terms may appear to make our Lagrangian pathological under the production of unphysical states (such as negative-energy ghosts), we note that this is in fact just a choice of gauge \cite{Cai:2014bda}. To make contact with the usual formulation of minimally coupled gravity, we impose the gauge-fixing condition $\chi^2-\psi^2 =6\alpha$ (the ``rapidity'' gauge), and hence parametrize the remaining degree of freedom by $\tanh(\frac{\phi}{\sqrt{6\alpha}}) = \frac{\psi}{\chi}$ in terms of a single scalar field\footnote{Note that this parametrisation does not fully capture all possible values of $\phi$. We may also write $\frac{\psi}{\chi} = \coth (\frac{\phi}{\sqrt{6\alpha}} )$. Depending on the form of $F$, the two parametrisations may not overlap, giving rise to distinct models \cite{Karamitsos:2019vor}.}. Noting that $d\psi = \chi d\phi$ and $d\chi = \psi d \phi$, we find that the Lagrangian can be recast in the following more familiar form:
\be 
\label{attractorL}
\L = \sqrt{-g} \left[\frac12R - \frac12\partial_\mu \phi \partial^\mu \phi  - F\Big(\tanh \frac{\phi}{\sqrt{6\alpha}}\Big) \right].  \ee 
It is worthwhile to note at this point the role played by $F$ in this construction. The function $\tanh$ has a range of $(-1,1)$ for real values of $\phi$, and is a monotonic function on the entire real line. Therefore, we consider functions $F$ which are differentiable  on this domain. Such functions will asymptote to constants for large values of $\phi$ when convoluted with $\tanh$, since $d(F \circ \tanh) = (dF \circ \tanh) \sech^2 $. Thus, if $F$ is differentiable on the domain, the convolution asymptotes to a constant and the ${\rm SO}(1,1)$ symmetry will be restored asymptotically. However, if we consider functions $F$ which are not differentiable everywhere, and in particular functions which have a pole in $(-1,1)$, the domain of the convolution will no longer be finite and the symmetry cannot be restored at large values of~$\phi$. Such a system was investigated in \cite{Dimopoulos:2016yep}, where it was shown that placing simple poles at the end points of the domain allowed for the reconstruction of familiar inflationary potentials. It is easy to see that setting $F=G \circ \arctanh$ allows the reconstruction of any function $G$ via such a process. This is possible exactly because $\arctanh$ is monotonic on the interval $[-1,1]$ with poles at the endpoints. We further note that our choice of gauge is qualitatively unimportant except that it imposes the restriction $\chi^2 > \psi^2$ (otherwise our system exhibits the wrong sign for gravitational couplings and will be unstable, leading to the production of ghosts).

The inflationary models captured by Eq.~\eqref{attractorL} are known as ``attractor models", in reference to their convergent predictions \cite{Kallosh:2013yoa}. If $F$ is chosen to be an even monomial function, the resulting models are often termed ``T-models'' due to the shape of their potential, which features plateaus away from the origin. Here, we must note that these plateaus arise as a result of the non-compact topology of the symmetry group we have employed. To show this, consider our scalar field $\mathbf{\phi}$ to live upon~$\mathbf{R}^n$. Upon gauge fixing such that the action reproduces the standard minimally coupled Einstein--Hilbert gravitational action, there remains a non-compact direction in the symmetry group of the matter field, since the gauge fixing must map the invariant function of the matter variables onto a positive, real constant 
${F}(\mathbf{\phi})=C$. 
Surfaces of constant $C$ are non-compact (as they form the symmetry group), yet are parametrized by members of $\mathbf{R}^n / \mathbf{R}_+ = {\rm SO}(n)$, which is a compact group. Therefore, in order to cover the entire symmetry group, we require an invertible map from a non-compact space onto a compact space, which must be a monotonic function with finite range yet infinite domain. Such functions must approach constants as their arguments tend to infinity, and hence necessarily form plateaus when taking on the role of a potential under any finite convolutions. As a contrasting case, if we had used $SO(2)$ as our symmetry group, the map would have been $\tan$, and hence our potentials would not be bounded from above. Thus, T-models are qualitatively generic for non-compact symmetry groups. 

We now consider a homogeneous and isotropic cosmology on the manifold $\mathcal{M}=\mathbf{R} \times S^3 $. Spatial slices have constant positive curvature, and correspond to $k>0$ Robertson--Walker geometries. In such a system, there is only one dynamical geometric degree of freedom, the scale factor $a = a(t)$. The line element in this case becomes
\be ds^2 = -dt^2 + a^2 \left[\frac{dr^2}{1-k r^2} +r^2 (d\theta^2 + \sin^2 d\theta^2) \right]. \ee
Thus, once we restrict ourselves to matter compatible with the space-time symmetries (i.e. homogeneity and isotropy), we are left with a familiar inflationary Lagrangian
\be 
\label{Lag} 
a^{-3}\L = 3 \bigg( \frac{\dot{a}^2}{a^2} - \frac{k}{ a^2}\bigg) - \bigg(\frac12\dot{\phi}^2 - V(\phi) \bigg), 
\ee
where $V(\phi)$ = $F \circ \tanh \frac{\phi}{\sqrt{6}}$. Throughout the rest of this work, we will perform our analysis for even monomials, i.e. $F(x) \propto x^{2n}$, e.g. $V(\phi)=\Lambda \tanh^2(\frac{\phi}{\sqrt{6\alpha}})$. 
A complete analysis of these models in the context of dynamical systems was recently completed \cite{Alho:2017opd}, in which the extended behaviour of the entire system was found to have unique endpoints under evolution, and the universality of such evolutions was established. 

\section{Background Dynamics}
\label{background}
In this section, we will follow the analysis of Ref.~\cite{Ellis:2015bag}, in which a scalar field was used to give rise to bouncing cosmologies as a counterpart to a decaying cosmological constant.  We will focus on the simplest potentials that can arise as a consequence of the broken $SO(1,1)$ symmetry. 
From our Lagrangian in Eq.~\eqref{Lag}, we obtain the familiar Friedmann equation, where $H = \dot a/a$:
\be H^2 = \frac13\rho  - \f{k}{a^2} \label{Friedmann}. \ee
It is immediately apparent that one cannot have $H=0$ (or equivalently $\dot{a}=0$) simultaneously with positive energy density and non-positive curvature $k \leq 0$. There exist a number of models which exhibit classical bounces through the use of ghost-matter ($\rho<0$) or phantom ($P<-\rho$) matter. Such matter may be introduced by fiat in order to remove the singularity \cite{Brown:2004cs} or as a result of a more fundamental physical process (such as the evolution of the fine-structure constant \cite{Barrow:2013uza}). However, in this work our goal, is to examine bounces in a classical context with matter which obeys the usual (weak) energy condition. To achieve this, we require $\Omega_k<0$. Thus, a bounce can only occur when $\rho=3 k/a_b^2$, where $a_b$ is the scale factor at the bounce (with orientation chosen such that $a>0$ at all times). At this point, we can expect either a bounce or a recollapse, depending on the sign of the second derivative of the scale factor (positive or negative, respectively).

We now examine how the conditions for a bounce are reflected in Raychaudhuri's equation, which is
\be \f{\ddot{a}}{a} = - \f{1}{6} \left( \rho + 3p \right). \ee 
Hence, a bounce occurs if $p < -\f{1}{3} \rho$. In terms of a barotropic perfect fluid, we require $w<-1/3$. The matter during a bounce will consist of a homogeneous, isotropic scalar field with a suitable potential. Since the equation of state depends upon the relationship between kinetic energy and potential energy of the scalar field, the same field can play multiple roles; indeed, this is one of the motivations for invoking such fields as inflatons. For a homogeneous, isotropic scalar field, the energy density is given by
\be \rho= \frac12\dot{\phi}^2 + V(\phi) \ee
and the pressure is
\be p = \frac12\dot{\phi}^2  - V(\phi). \ee
As a result, $w$ is given by
\be w = \f{p}{\rho} =  \f{\dot{\phi}^2/2 - V(\phi)}{\dot{\phi}^2/2+V(\phi)}. \ee 
We see that $w=1$ corresponds to kinetic domination and $w=-1$ corresponds to potential domination. Therefore, we require $V(\phi) > \dot{\phi}^2$ in order to have $w<-1/3$. 

We now consider a general model of a potential featuring a plateau, given by $V(\phi) = \Lambda u(\phi)$ where the maximum value of the function $u(\phi)$ is 1 as $\phi \to \pm \infty$ (without loss of generality). This potential generalizes the usual T-models, and can arise from functions $F$ which do not necessarily have poles.
% in $[-1,1]$
As a result, potential domination means that any bounce must occur when
\begin{equation}
\label{lambdaconstr}
\rho \leq \frac{ \Lambda}{2}.
%\rho \leq \frac{3 \Lambda}{2}.
\end{equation}
Note that if our field is to both provide the dynamics of the bounce and play the role of inflaton, this constraint indicates that the bounce must occur far from the quantum gravity scale. This is because the energy scale of inflation is of the order of $10^{16}$ GeV, whereas the Planck scale (where we expect quantum gravity effects to become important) is around $10^{19}$ GeV. In such cases, our universe will bounce from contracting to expanding without ever approaching the quantum gravity scale. In turn, we find that the minimal radius of curvature at which a bounce can occur is $l_m = \sqrt{\frac{2}{  \Lambda}}$.
From the Planck data, we find that this is far from the Planck length for $|\Omega_k|$ below the measured upper bound as well as above the lower bound of being indistinguishable from a super-horizon fluctuation. Hence, the universe can indeed undergo a bounce wherein both the matter and geometrical sectors are far from the quantum gravity regime. 

We can use the condition \eqref{lambdaconstr} together with the Friedmann equation \eqref{Friedmann} to show that any value of $w$ between $-1$ and $1$ can be realised by imposing the following condition at the bounce point ($a=a_b$):
\be V(\phi) = \f{3 k}{a_b^2} (1-w)  \quad\quad \dot{\phi}^2 = \frac{6k}{a_b^2} (1+w)  \label{bouncecond} \ee 
Recall that if $w>-1/3$, this is a point of recollapse rather than a bounce. Thus, all we must strictly require is that $V(\phi)$ has a range which contains this value. In terms of the T-model potentials described by $V(\phi) = \Lambda \tanh^{2n} \left( \frac{\phi}{\sqrt{6\alpha}} \right)$, this is achieved by
\be \phi > \sqrt{6\alpha} \arctanh \left[ \left(\frac{2  k}{  a_b^2}\right)^{\frac{1}{2n}}\right]. \ee
Note that this a necessary but not sufficient condition for a bounce. In order for the bounce to occur, we also require that the potential energy is larger than twice the kinetic energy. Furthermore, we must stress that we are considering the dynamics of a universe which is contracting: as such, the usual Hubble friction term~$3H\dot\phi$ gives rise to \emph{anti-friction} when $H<0$, serving to increase the kinetic energy instead. Therefore, when the inflaton reaches the plateau in a contracting branch of the Friedmann equation, it is near a minimum of kinetic energy density on the solution (and hence a maximum of $w$). If this is not sufficient to cause a bounce (i.e. if we have $w>-1/3$), then the inflaton will accelerate, the kinetic component of energy density will dominate the evolution, and the system will reach a singularity. During the final preparation of this manuscript we were made aware of a similar discussion in \cite{Matsui:2019ygj}, which parallels much of the background dynamics. 

An important issue arises when considering models beyond the homogeneous and isotropic background dynamics. The de Sitter expansion is an attractor in the space of inhomogeneous and isotropic cosmologies \cite{Sloan:2015bha,Sloan:2016nnx}. As such we should be unsurprised to see after inflation a universe that is homogeneous and isotropic; this was one of the key motivations behind inflationary models in the first instance. However during a collapsing phase the opposite is true. Inhomogeneities grow to a point at which their effects may overwhelm the background dynamics and thus avoid the bounce. Therefore we must address the issue of whether such a bounce can be physically realized from initial conditions set far before the bounce, such as at the onset of collapse. 

In the model we present we have restricted the matter content to be that of the scalar field alone, as this is the dominant contribution to the stress-energy tensor at high energy scales. If we examine the pre-bounce phase of such a solution we find that as the universe expands the scalar field finds the minimum of its potential, and in oscillating around it appears to have an effective equation of state similar to that of dust, with $w=0$ or even faster fall-off, with averaged $w>0$. See the appendix for more details of this. Hence at large scale factor this model becomes dominated by the curvature term and reaches a finite extent. Further to the past this model is again in an expanding phase and becoming homogeneous. Therefore setting initial conditions far in the past would not necessarily be an impediment to the universe undergoing a bounce. A universe which begins at a singularity with some inhomogeneity, expands and recollapses can have smaller inhomogeneity at the bounce point. See section \ref{numerics} and figure (\ref{BounceFromNothing} therein for a numerical simulation demonstrating this possibility.

In a more physically motivated model we should also consider the introduction of a (small) cosmological constant to make contact with late time observations in our own universe, and consider the case where the inflaton decays into other matter. The background dynamics of such a model can begin from a universe of infinite size at $t\rightarrow -\infty$, however this is also not necessarily the case. Such universes can also begin at singularities, or have repeated bounce and recollapses. As an example, consider a universe at a bounce point which has the scalar field low enough on its potential, and rolling down the potential, such that there is not a lot of inflation. Such a universe may never become dominated by the cosmological constant, but it would recollapse to a singularity. The time-reverse of such a model is a universe that begins at a singularity, and recollapses to a bounce then expands through inflation to a de Sitter universe. 

Finally we note that in order to match with observations of our universe it is not necessary that the entire universe bounce, but that at least a single patch of approximately size corresponding to the Hubble radius during inflation bounces. As such we can consider the scenario discussed in \cite{Ellis:2015bag} in which initial conditions are set far in the past for an inhomogeneous universe. If the inhomogeneities are randomly distributed in a large universe, the majority of any section of a spatial neighborhood may be dominated by their inhomogeneities during collapse. This will lead to these regions becoming singular. However some sections will have sufficiently small inhomogeneities that do not spoil the bounce, and thus expand again, and undergo inflation. This inflationary period will push any singular regions far beyond the Hubble horizon of the expanding universe after the bounce, possibly beyond the asymptotic de Sitter horizon also. \footnote{If the universe undergoes eternal inflation during this period, this one patch can seed a large, even potentially infinite, number of patches which later may recollapse with their own inhomogeneities. Many of these patches will again become singular, but with a large number some of them will by chance be sufficiently homogeneous that they can again bounce and the cycle continues.}  Thus as observers we should not be surprised to find ourselves in the region which underwent a bounce any more than we should be surprised to find ourselves in a universe where the cosmological constant is compatible with our existence. 

\section{Phase Space}
\label{Phasespace}

The Hamiltonian treatment for our model follows directly from the Lagrangian in Eq.~\eqref{Lag}, with canonical structure inherited from the conformal theory in Eq.~\eqref{conf1}. For clarity of exposition, we will work with the volume $\nu=a^3$ instead of the scale factor throughout this section. The primary reason for this is to make our algebra considerably more transparent, since the canonical conjugate to the $\nu$ is the Hubble parameter $H$, i.e. the Poisson bracket is given by $\{\nu,H\}=\frac{1}{2}$. Our phase space is thus parametrised by $\{\nu, H; \phi,\p\}$, where $P_{\phi}$ is the conjugate momentum. Our solutions are further determined by the value of the curvature $\Omega_k \equiv - k/\rho_c$, defined at the bounce (as opposed to today, as usual).

\emph{A priori}, we have a 5 dimensional space to consider. We then wish to find the space $\Gamma$ consisting of physically indistinguishable solutions. We find that the Hamiltonian $\H$ is (using the symbol $\equiv$ to note the constraint):
\be 
\label{hamiltonianconstraint}
\H = -  3H^2 \nu  + \frac{\p^2}{2\nu} + \nu V + \Omega_k \nu^{1/3} \equiv 0. \ee
Here is is useful to note that the Hamiltonian constraint in Eq.~\eqref{hamiltonianconstraint} and the Friedmann equation~\eqref{Friedmann} are equivalent. We therefore have a four-dimensional phase space with a single constraint (the diffeomorphism constraint that usually arises in the 3+1 decomposition of GR is automatically satisfied by the homogeneity of our system). Thus, we should expect to see a two-dimensional space of solutions to our systems generated by the Hamiltonian flow. The symplectic structure is given by the symplectic form 
\be \omega = 2  d\nu \wedge dH + d\phi \wedge d\p.\ee
From this, we find that our familiar Hamiltonian vector field on phase space now includes a term induced by the presence of curvature:
\be X_\H =  6 \nu H  \f{\partial}{\partial \nu} -  6  \left(\frac{\p^2}{\nu^2} + \frac{2 \Omega_k}{3\nu^{2/3}}\right) \frac{\partial}{\partial H} + \frac{\p}{\nu} \frac{\partial}{\partial \phi} - \nu V' \frac{\partial}{\partial \p}. \ee
Since we are in a closed universe ($\Omega_k <0$), the coefficient of $\frac{\partial}{\partial H}$ is not necessarily positive. In other terms, the Hubble parameter is not monotonically non-increasing on all solutions. This means that we cannot choose a fixed Hubble slice to parametrize the space of solutions to our model since there will be solutions which cross a given value of the Hubble multiple times (e.g. in both expanding and recollapsing phases). Moreover, we cannot necessarily count solutions at the bounce/recollapse point. This is because $\Lambda$-dominated late-time solutions will not recollapse, and small-scale kinetically dominated solutions will not bounce. However, if we restrict ourselves to subspaces of the set of physical solutions, we can often choose our subsets to have well-defined properties. As an example, consider the space of solutions which undergo a bounce. These necessarily satisfy $H=0$ and $w<-1/3$ simultaneously. Thus, we can count such solutions by constructing a measure on a subset of the full phase space that intersect these conditions. These subspaces may yet be non-compact; in this particular instance, the intersection of the Hamiltonian constraint with the $H=0$ surface is non-compact, since any choice $\nu \in \mathbf{R}_+$ can always satisfy the constraint for a fixed $\Omega_k$, simply by scaling $\p$ accordingly. 

The solution to this issue is familiar in the flat case \cite{Corichi:2013kua}, and involves the restriction of the analysis to physical degrees of freedom rather than phase-space variables. Solutions related by rescaling of both $\p$ and $\nu$ such that their ratio remains fixed are dynamically related. Indeed, it is this ratio which constitutes the physical parameter $\dot{\phi}$ (both the equations of motion for $H$ and~$\phi$ are written only in terms of $\dot{\phi}$ not $\nu$ nor~$\p$ alone in the flat case). When introducing curvature, it would appear that this is no longer the case because $\nu$ contributes directly to $\dot{H}$ through the curvature. However, we note again that $\Omega_k$ is only determined up to an overall scaling in an intrinsic description of physics; a simultaneous rescaling of $\Omega_k$, $\nu$ and $\p$ can be performed such that there is no physical distinction to a cosmologist working within such a universe. 

It may seem strange to the reader that we are allowing a change in the constants of our system under the action of this rescaling symmetry. We note that any observer within this system would only be able to infer $\Omega_k$ through observations of the physical degrees of freedom of their universe (such as for example, $\dot{H}$), which are unaffected by this change. As such, a cosmologist living in this universe could only determine physics up to a choice of fixing this symmetry. In fact, we have taken this into account by fixing $\sum \Omega_i=1$. Such a change is generated by the action on the extended space of both phase space and the constants of motion by
\be \mathcal{G} = \p \frac{\partial}{\partial \p} + \nu \frac{\partial}{\partial \nu} + \Omega_k ^{2/3} \frac{\partial}{\partial \Omega_k}. \ee
In the above expression, $\mathcal{G}$ is the adaptation of the minimal coupling symmetry explored in \cite{Sloan:2014jra} to a curved cosmology. It commutes with the Hamiltonian constraint and preserves the 
Hamiltonian flow,
and thus does not affect the dynamics of the intrinsic system. Note that $\cal G$ is not a symplectomorphism, but rather a ``scaled" symplectomorphism: $\mathbf{L}_G \omega \propto \omega$, where $\mathbf{L}$ denotes the Lie derivative. We can employ the generator $\cal G$ in order to project our space of solutions, essentially gauge fixing $\Omega_k$ to any given value. Equivalently, our space of bouncing solutions $\Gamma_b$ is determined at the point of bounce by the triple $\{\phi_b,\nu_b,\Omega_{k}\}$, modulo the action of $\cal G$ which acts as an equivalence relation between solutions 
\be \{\phi_b,\nu_b,\Omega_{k}\} \sim \{\phi_b,\beta \nu_b, \beta^{2/3} \Omega_{k}\} \quad \forall \beta \in \mathbf{R}_+ \; . \ee
Thus, we can choose a member of each equivalence class by setting $\Omega_k=-1$, for example. In order to evaluate our space of solutions, however, it is more convenient to work with the invariants of $\cal G$; the first is the ratio $\p/\nu = \dot{\phi}$, and we choose the second to be $\xi=\Omega_k^{2/3}/\nu$.  In these terms, the symplectic structure pulled back to the $H=0$ surface (denoted by $ \overleftarrow{\omega}$) becomes
\be \overleftarrow{\omega} = \frac{2\xi-3V}{2\sqrt{\xi-V}} \frac{\Omega_k^{3/2}}{\xi^{5/2}} d\phi \wedge d\xi .\ee
We note that the orientation of $\overleftarrow{\omega}$ can flip depending on the sign of $2\xi-3V$. When positive, this measures solutions which are at the point of recollapse. Conversely, when negative, it measures solutions that bounce. Although $\Omega_k$ appears in the expression of $\overleftarrow{\omega}$, this is merely an overall choice of scale (due to the fact that $\cal G$ is a scaled symplectomorphism). It will not contribute to any calculation of the fraction of solutions that have a certain property. 

We have thus identified the space of bouncing solutions, $\Gamma_b$, to be the subspace defined by the intersection
$\mathcal{H}^{-1} (0)  \cap (V< F < 3V/2)$.
Its symplectic structure is non-degenerate (due to being a two-form on a two dimensional space of bouncing solutions $\Gamma_b$), and hence can act as a natural measure on the space, derived from the Liouville measure. Any other measure can be obtained by taking $\overleftarrow{\omega}$ and multiplying by some function of $\xi$ and $\phi$. This space is still non-compact for potentials which feature a plateau; this is unsurprising as it reflects the fact that the plateau of the potential in the infinite limit restores the symmetry of the system and the physical differences between solutions become indistinguishable for large $|\phi_b|$. Ratios of the measure on the space of bouncing solutions can be calculated by introducing a cut-off in $\phi$ beyond which we identify solutions due to the stochastic effects overwhelming the classical dynamics. There is a pole of order $\xi^2$ about the second symmetric point where $V=0$, which can also be regularized either by identifying solutions with $|\phi|<\epsilon$ for some $\epsilon>0$ or equivalently assuming $V$ has a small non-zero minimum. Evaluating this measure, we find that our system is dominated by solutions that bounce with the the particle on the plateau, giving the right conditions for a long period of slow-roll inflation to follow. This result holds qualitatively for all plateau models, since the set of $\phi$ that satisfy $V(\phi)<B$ for some~$B$ becomes the entire real line when $B$ exceeds the height of the plateau. In a similar manner to the choice of $F$ overcoming the existence of the plateau, the choice of function used with the volume form could overturn this result. However, such a function would have to disfavour values of $V$ which allow the particle to reach the plateau, and thus would also require poles.

\section{Approximate Analytic Solutions}
\label{analytic}
To provide a clear explanation of why we observe bounces in such models, let us consider the cosmological system in two limiting cases of energy in a collapsing universe. In the low energy limit $\rho \ll \Lambda$, the inflaton is low on the potential. The dynamics of the field are those of a harmonic oscillator with a small anti-friction term (a greatly under-damped system). As such, $w$ oscillates between the extreme values of $\pm 1$, and so on average (across a cycle), we can model the system with an approximately constant equation of state with barotropic parameter $\overline{w}=0$. As the energy density increases, this approximation becomes invalid and $w$ decreases; see Appendix \ref{hyptan} for details. Thus, our dynamics is akin to that of a closed universe with dust as the dominant matter component, and the Friedmann equation becomes
\be H^2 = \frac{1}{3} \left(\frac{\Omega_d}{a^3} + \frac{\Omega_k}{a^2}\right), \ee
where $\Omega_d$ is a constant. We reiterate at this point that since we are dealing with a closed universe,~$\Omega_k$ is negative. Such models have simple parametric solutions in conformal time $dt=ad\eta$, given by:
\begin{align}
a &\propto  1-\cos   \eta,
\\
 t &\propto  \eta-\sin  \eta.
\end{align}
Since we are in the regime where the scalar field oscillations are small (and hence both the harmonic oscillator approximation and the dust-like behaviour are valid), these are valid in our system around the points of recollapse ($\eta \approx \pi$). There is a well-known redundancy in our description of these models: rescaling the constants $\Omega_i$ by an appropriate power law relation yields equivalent dynamics, since each constant now carries a dimensionful scale. The general form of this symmetry gives rise to the attractor solutions in cosmology by imposing that the Liouville measure is preserved (see Refs.~\cite{Corichi:2013kua,Sloan:2015bha,Sloan:2016nnx}). 

In this section, we will deviate from the convention of fixing the sum of the $\Omega_i$ to unity. Our goal here is to discuss an approximate analytic solution to the equations of motion, and the exact equations of motion contain the evolution of the inflaton, for which the equation of state is dynamic. In essence, we shall show that the dual role played by the inflaton (which sometimes acts as cosmological constant and other times as a stiff fluid) can be mimicked by transfer between the different $\Omega$. Due to the distinct scaling of these (as required in order to preserve the dynamics), a transfer between two components will change their sum. Thus, in order to obey the convention, each $\Omega$ would have to change. Instead, we opt to keep $\Omega_k$ fixed during evolution, since there is no transfer between it and the matter components. This is because $\Omega_k$ arises from the geometry of the spatial slice which we do not couple to matter.   

We are using the harmonic oscillator approximation, which is valid because we are considering the simplest form of the function, given by $F(x)=x^2$ . For higher powers, and appropriate rescaling of the width, the potential approaches that of a square well, with potential given by
\be V = 
\begin{cases} 
\Lambda \qquad |\phi| > \theta/2 \\ 
0  \qquad |\phi| < \theta/2  ,  \end{cases} \ee
where $\pm \theta/2$ are the approximate endpoints of the plateau.
We first consider the behaviour when the particle is in the well ($\rho < \Lambda$). In this case, our dynamics is given by
\be  H^2 = \frac{1}{3} \left( \frac{\Omega_o}{a^6} + \frac{\Omega_k}{a^2} \right). \label{InWell} \ee
Again, this equation can be solved exactly in parametric form by considering conformal time wherein we find 
\be a^2= \sqrt{-\frac{\Omega_o}{\Omega_k}} \cos \left(\sqrt{-\frac{8  \Omega_k}{3}} (\eta-\eta_r)\right),\ee
where $\eta_r$ is the conformal time at which the universe begins to recollapse, i.e. the scale factor reaches its maximal value. We can then recover $t$ through an elliptic integral.

When we consider the case where $\rho>\Lambda$, the particle will not oscillate, but will behave as if it is on a flat potential. This can be further split into the constant term (the height of the potential) which behaves like a cosmological constant, and the kinetic term, which evolves like a massless scalar field.
Thus, the Friedmann equation is well approximated by 
\be H^2 = \frac{1}{3} \left(\Lambda + \frac{\Omega_s}{a^6} + \frac{\Omega_k}{a^2} \right). \label{SWLO} \ee
If $\Omega_s=0$ in such models, we recover the closed de Sitter solution given by 
\be
 a=\sqrt{-\frac{\Omega_k}{\Lambda}} \cosh \left[ \sqrt{\frac{ \Lambda}{3}} (t-t_o) \right].
\ee
In the general case in which the kinetic component scalar field is non-zero, we find that a bounce occurs if $\Omega_s$ is less than a critical value~$\Omega_c$. To determine $\Omega_c$, consider that the condition for a bounce in Eq.~\eqref{SWLO} becomes a cubic equation in $b=a^2$:
\be b^3 +\frac{\Omega_k}{\Lambda} b^2 + \frac{\Omega_s}{\Lambda} = 0, \ee
and the critical condition is that this occurs at a local minimum of $H$ over $a$. To see why this is true, consider the roots of our cubic equation for some value of $\Omega_s$. Since both $\Lambda$ and $\Omega_s$ are positive, one root is negative and thus not physical. Thus, we consider either of the two positive roots. If this is not a local minimum of the cubic function, we can increase $\Omega_s$, which increases $H$ and continue to do this until the cubic no longer has a root. This happens when the root is at the minimum of the cubic (and hence a repeated root):
\be 3b^2 + 2\frac{\Omega_c}{\Lambda} b=0 .\ee
We solve this for $b$, and reinsert into the cubic equation to find the critical value $\Omega_c$, which is 
\be \Omega_c = -\frac{4 \Omega_k^3}{27 \Lambda^2} \label{Ocrit} .\ee
These solutions, well away from the bounce point, asymptote to the de Sitter solutions above. However, as the solution evolves from the bounce point, Hubble friction will eventually slow the scalar field to a halt.\footnote{For the moment, we ignore the effect of quantum fluctuations. We discuss this later on.} Subsequently, it will turn around\footnote{A slight tilt in the plateau is assumed.} and, after a period of slow-roll inflation, fall down the potential removing the effective cosmological constant term, allowing the expanding universe to recollapse. We note that if a period of constant slow-roll or ultra-slow roll occurs due to the flatness of the potential, it is going to be necessarily transient \cite{usr,Morse:2018kda,Lin:2019fcz}.

To establish when general solutions will bounce, we consider the motion of the inflaton field itself. Consider the point at which the inflaton hits the wall. If the kinetic energy ($\frac{\dot{\phi}^2}{2}=\frac{\Omega_o}{a^6}$) is less than the height of the potential $\Lambda$, the field will reflect and continue, and we can repeat our analysis at the next interaction. If the kinetic energy is greater than the height of the wall, it will be reduced by $\Lambda$ as it traverses the wall, and hence $\Omega_s = \Omega_o - \Lambda a^6$. To obtain an analytic solution for such a scenario, we will neglect the contributions of the kinetic terms of the scalar field when on the plateau and the potential terms when in the valley. We also note that conservation of the scalar field energy density during the transition ensures the continuity of the Hubble parameter. 

Between the reflections, we can find the behaviour of the inflaton. Using $\dot{\phi}^2/2 = \Omega_o/a^6$, we can write the equation of motion as follows in terms of the scale factor:
\be \frac{d\phi}{da} = \frac{\dot{\phi}}{\dot{a}} = \frac{\dot{\phi}}{Ha} = \frac{1}{\Omega} \sqrt{\frac{6\Omega_o}{\Omega_o+a^4 \Omega_k} }  . \ee
This can be exactly integrated and inverted to give the scale factor $a$ in terms of $\phi$:
\be a^2 = \sqrt{-\frac{\Omega_o}{\Omega_k}} \sech\left[\sqrt{\frac{2}{3}} (\phi-\phi_o)\right] .\ee
From this we can express the kinetic energy in terms of $\phi$ as
\be  \frac12\dot\phi^2 = \sqrt{-\frac{\Omega_k^3}{\Omega_o}} \cosh^3\left[\sqrt{\frac{2}{3}} (\phi-\phi_o)\right]. \ee
We can therefore consider the following scenario: if the well is sufficiently narrow, the kinetic energy between reflections cannot increase by $\Omega_c$ or more, and hence the universe must bounce. The limiting case for a reflection happens when the kinetic energy is equal to the height of the potential. Therefore, we consider a reflection at $\phi=-\theta/2$ with kinetic energy $\Lambda$. For this to occur, we require that $\Lambda > -\Omega_k^{3/2}/\sqrt{\Omega_o}$ (otherwise the kinetic energy is always sufficient to escape the well, and our cosmology is always described by Eq.~\eqref{SWLO} and asymptotes to the $k>0$ de Sitter expansion). This fixes 
\be \phi_o = -\frac{\theta}{2} - \sqrt{\frac{3}{2}} \arccosh \left( \frac{\Omega_o^{1/6} \Lambda^{1/3}}{\sqrt{-\Omega_k}} \right). \ee
Note that the term inside the inverse hyperbolic cosine function is composed of a dimensionless combination of our constants. The particle next encounters the wall of the potential at $\phi=\theta/2$. If by this stage the kinetic energy has increased by less than $\Omega_c$, the remaining kinetic energy after the particle leaves the potential well will be sufficiently small that the universe will bounce. The amount of kinetic energy gained is a function of the width $\theta$ of the potential. Therefore, we can choose $\theta$ to ensure that the amount of kinetic energy gained is always less than this. This is achieved by:
\be \theta = \sqrt{\frac{3}{2}} \left[ \arccosh \left( \sqrt{-\frac{\Omega_o}{\Omega_k^3 }}\Lambda +4\frac{\sqrt{-\Omega_k^3 \Omega_o}}{27 \Lambda^2} \right)^{1/3}  - \arccosh \left( \sqrt{-\frac{\Omega_o}{\Omega_k^3 }}\Lambda \right)^{1/3} \right]. \ee
Thus we can, in principle, ensure that a given branch of the homogeneous, isotropic, and closed cosmology will always bounce, provided we choose the correct potential. The exact solution given here is derived in the case of the square well, which is the most analytically tractable situation. However, we see from Eq.~\eqref{Ocrit} that once the potential is well approximated by a plateau, a bounce will occur as long as the kinetic energy of the inflaton is sufficiently small. Therefore, for any of the generic T-model potentials, there will always exist in phase-space a set of conditions leading to a bouncing universe.\footnote{Unless, quantum fluctuations
come to dominate the variation of the field. However, in the next section we show that, for realistic T-model inflation, turnaround typically occurs before quantum fluctuations become important.} 

Beyond the bounce, our analytic approximation will break down. As the Hubble parameter is now positive, the system will encounter friction, and the scalar field will slow. Since the true potential is never actually flat, the combination of this friction and the gradient of the potential will be sufficient to halt the inflaton and reverse its motion, beginning a long period of slow roll back down the potential. To make this apparent, consider the behaviour of the scalar field on the plateau: for low kinetic energies, the dominant contribution to the Hubble parameter will be the potential, and hence we find that the velocity of the field follows
\be \dot{\phi} \propto \exp(-3 \Lambda t). \ee
As a result, the field will come to a halt at a finite value of $\phi$ after an infinite interval in time. However, since the true potential has a gradient, this behaviour cannot happen. In fact, the true dynamics will diverge from that of the square well when the Hubble friction term becomes of the same order as the gradient of the potential. There is always a point at which the field turns around, and the square well cannot capture this behaviour.

One way in which we can alleviate this problem is compactify the scalar field on an $S^{1}$ or set a reflection at some $\phi_r$ to represent the point at which the scalar field would turn around. We will focus our discussion on the latter case, this can recreate a phenomenologically interesting double-bouncing solution. This is behaviour is quantitatively distinct from the one which we would expect from a scalar field on a T-model potential. However, it encompasses the relevant qualitative features, namely that the field encounters the valley again after the bounce. To facilitate the analytic calculation of perturbations on this background, we express the evolution of the scale factor in terms of conformal time. We can explicitly construct a double-bouncing cosmological model using Eq.~\eqref{SWLO} during the bounces and Eq.~\eqref{InWell} for recollapse. Such solutions are continuous in $a$ and $H$, this condition determines the values $\eta_1$ and $\eta_2$ of conformal time at which we connect the descriptions; \emph{ab initio}, these are free parameters of our double-bouncing model, but will be constrained by the cosmological parameters.  We find that the evolution of the scalar factor is
\begin{figure}[htbp]
 {\includegraphics[width=0.48\linewidth]{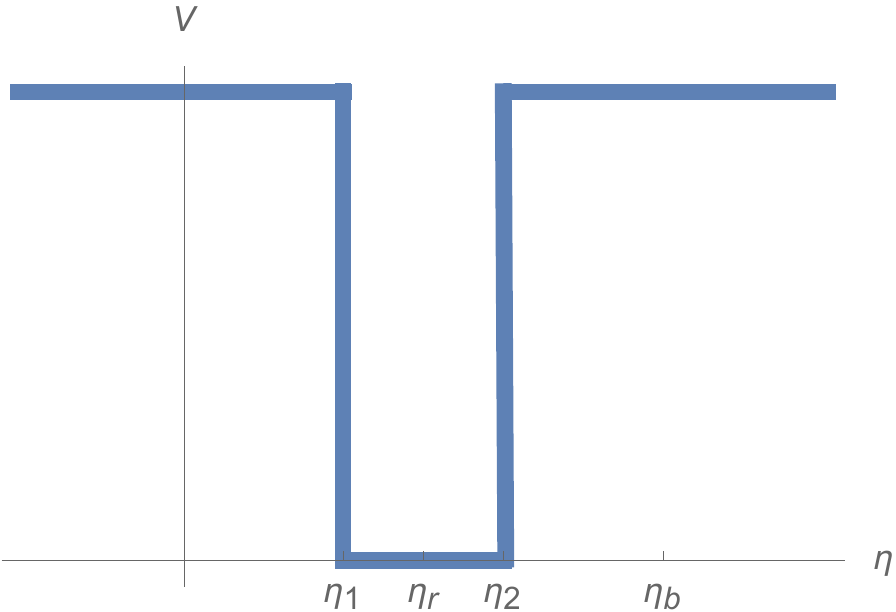}}
\caption{The double-bouncing cosmological model under discussion in this section. The evolution of the potential is shown with respect to conformal time. The first bounce occurs at $\eta = 0$ and the recollapse occurs at $\eta = \eta_r$.}
\label{DoubleBounceGraphic}
\end{figure}
\be
\label{db} 
 a = 
\begin{cases} 
\sqrt{-\frac{\Omega_k}{\Lambda}} \sec(\omega \eta)  &  \eta < \eta_1, \\ 
                                   \left( \sqrt{-\frac{\Omega_s}{\Omega_k}} \cos [2 \omega (\eta-\eta_r)] \right)^{1/2} &  \eta_1 < \eta < \eta_2, \\
                                    \sqrt{-\frac{\Omega_k}{\Lambda}} \sec[\omega (\eta-\eta_b)] & \eta > \eta_2, 
\end{cases}
\ee
where $\omega=\sqrt{- \frac{  \Omega_k} {3}}$.

 Without loss of generality, we choose the first bounce to occur at $\eta=0$, recollapse at $\eta=\eta_r$ and second bounce at $\eta=\eta_b$. The transitions between the distinct descriptions occur at $\eta=\eta_1,\eta_2$. From this result, we immediately identify the value of the scale factor at the bounce as $a_b = \sqrt{-\frac{\Omega_k}{\Lambda}}$ and at recollapse $a_r = (-\frac{\Omega_s}{\Omega_k})^{1/4}$. Conservation of energy at transitions between plateau and valley sets the scale factor here as $\Omega_s=\Lambda a_t^6$. Thus, we find that these value of the scale factor at these times are related to one another. Indeed, we can express the scale factor at recollapse in terms of that at the bounce and that of the transitions:
 \be a_r = \left(-\frac{\Omega_s}{\Omega_k}\right)^{1/4} = \left(-\frac{\Lambda a_t^6}{\Omega_k}\right)^{1/4} = \sqrt{\frac{a_t^3}{a_b}}. \ee
 Thus, if we parametrize our space of double-bouncing cosmologies by the ratio of scale factor at recollapse to that at the bounce, $\gamma = a_r/a_b$, then our system is completely specified. Our approximations hold most closely in the symmetric solution where the universe recollapses with the inflaton at the center of the square well (where all the energy is kinetic, and thus $w=1$). As a result, all energy lost to Hubble friction in the expanding phase is regained during the collapsing phase, and the inflaton has sufficient energy to reach the plateau when it encounters the wall. Thus, we can fix the parameters of  \eqref{db}: $\eta_1 = \frac{1}{\sqrt{k}} \arccos(\gamma^{-\frac{2}{3}}), \eta_r = 3\eta_1/2, \eta_2 = 2\eta_1, \eta_b = 3\eta_1$. The behaviour of this system is shown in Fig.~\ref{DoubleBounceGraphic}.
 
We end this section by noting that the conditions which relate the width of the potential to the parameters $\Omega_k$, $\Omega_o$ and $\Lambda$ guarantee that the cosmology will undergo a bounce. However, $\Omega_o$ is dynamically generated; unlike $\Omega_k$ and $\Lambda$, it changes when the particle encounters the well. It will not necessarily take the same value each time the particle falls into (or leaves) the well, since it depends on the amount of Hubble friction (or anti-friction in contracting cases) that the scalar field encounters during its evolution. In the particular example of the square well, we find that $\Omega_o = a^6 (\frac12\dot{\phi}^2 + \Lambda)$ at the point at which the particle reaches the well if we allow the inflaton to reach it after a bounce has occurred. Therefore a fixed potential will not guarantee that the resulting cosmology is never singular, although it may guarantee a bounce for a given combination of $\Omega_k, \Lambda$ and $\Omega_s$.

\section{Numerical Evolutions}
\label{numerics}
\begin{figure}[htbp]
\subfloat[]{\includegraphics[width=0.48\linewidth]{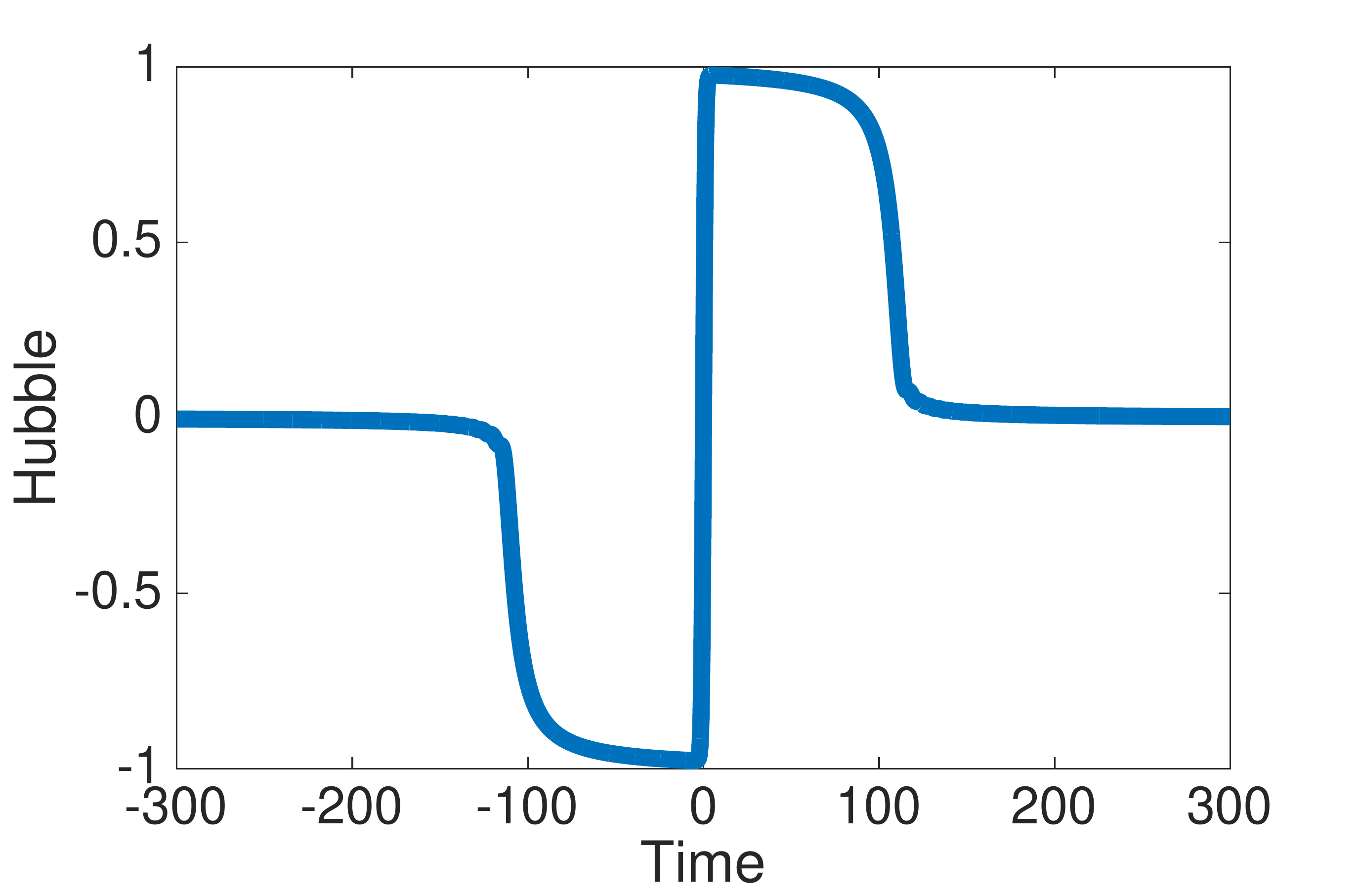}}
\subfloat[]{\includegraphics[width=0.48\linewidth]{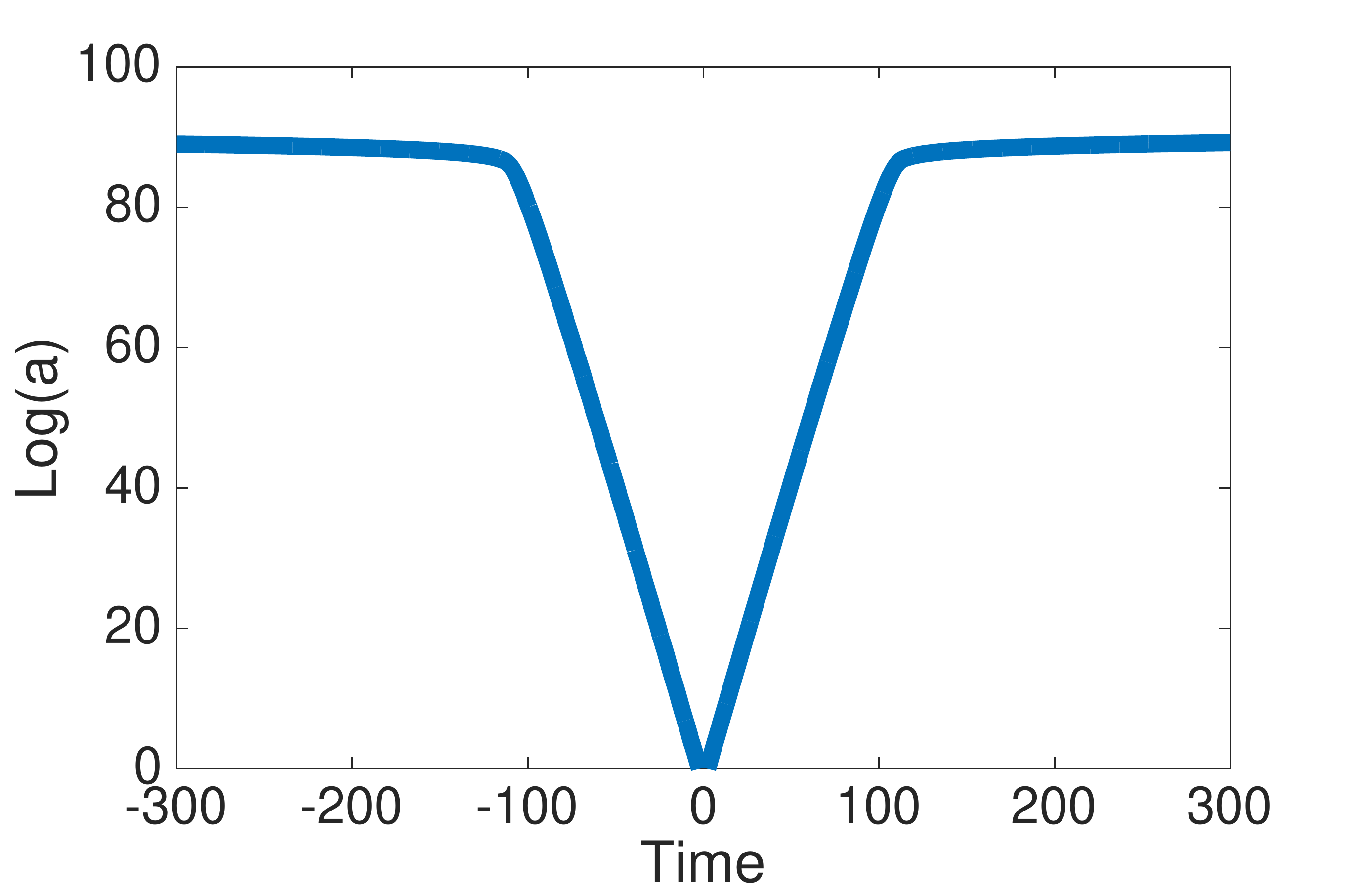}}
\caption{Hubble (a) and log of scale factor (b) against time for a solution with a bounce at $t=0$. We see a typical evolution with singularities in both the future and past with a bounce between them. This displays the behaviour of a simple bounce with a large number of $e$-folds of inflation/deflation on either side. This is a toy model with $\alpha=1, \, \mu=1,  \, a=0.224, \, \Omega_k = -0.05$, and $w=-0.9$ at the bounce.}
\label{H}
\end{figure}

In Fig.~\ref{H}, we show a typical bouncing solution chosen with $w=-0.9$ at the bounce point. Such solutions can undergo a long period of inflation whilst the inflaton is on the plateau of the potential, and produce a spectral tilt and scalar-to-tensor ratio which agree with observations. In such solutions, the scalar field is high on the potential at the bounce, with the inflationary/deflationary periods happening with the scalar field on the same plateau. There is an asymmetry between the expanding and contracting phases due to the velocity of the scalar field at the bounce; as the inflaton approaches the plateau its kinetic energy drops, pushing $w$ towards $-1$. However, if there is sufficient energy to reach the plateau, the anti-friction will again reaccelerate the inflaton. 

\begin{figure}[htbp]
\begin{center}
\subfloat[]{\includegraphics[width=0.48\linewidth]{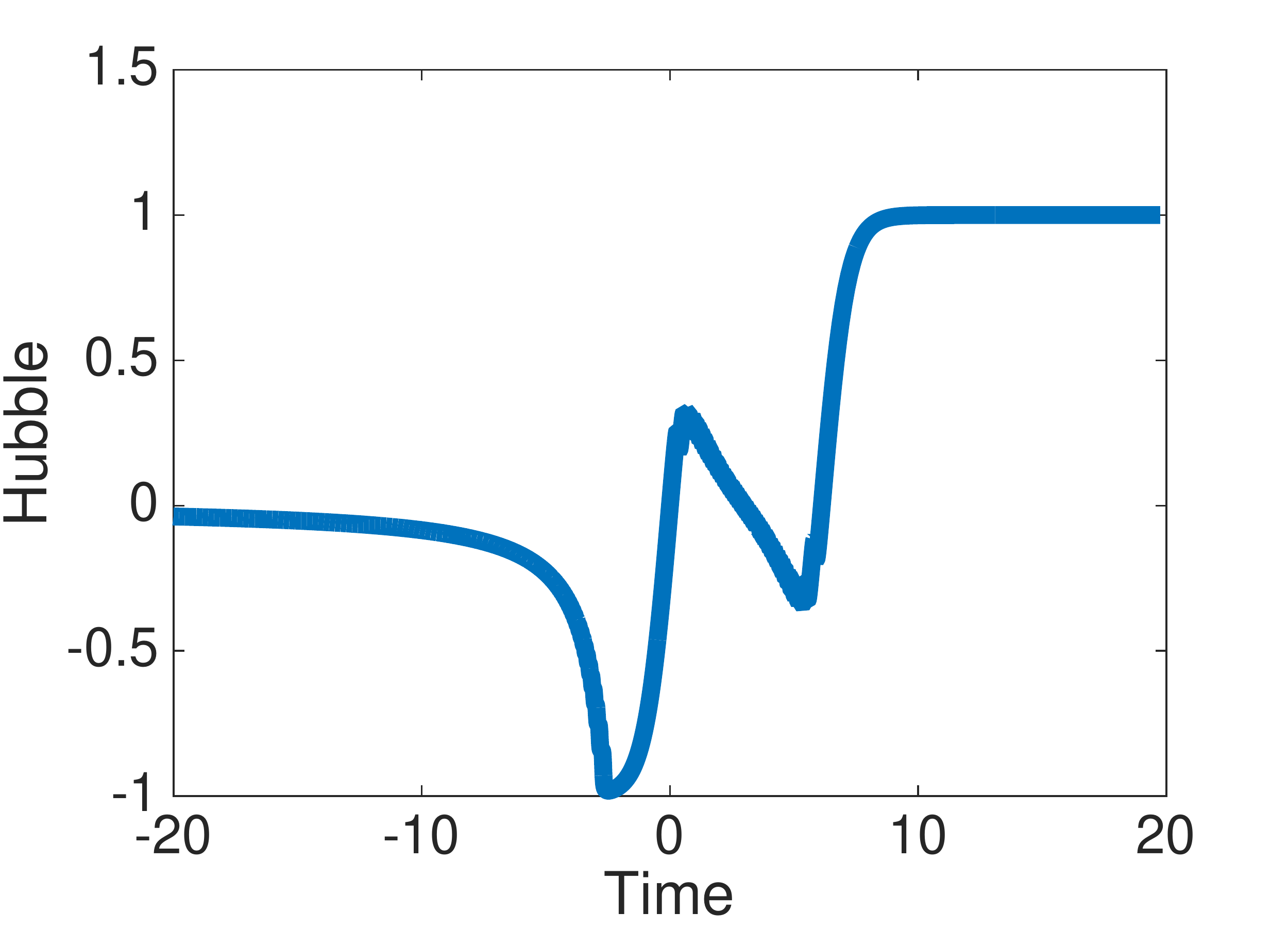}}
\subfloat[]{\includegraphics[width=0.48\linewidth]{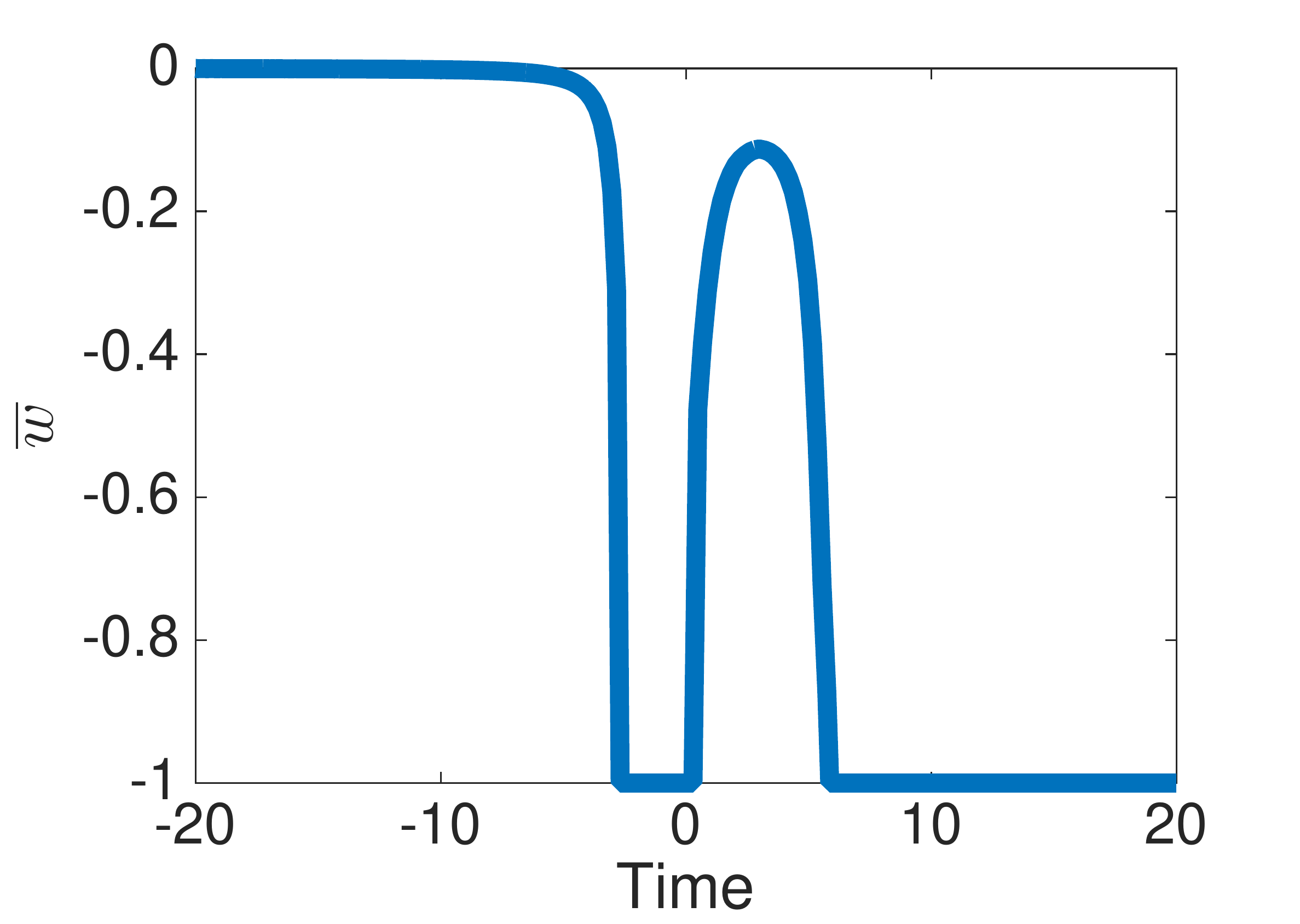}}
\caption{On the left, the evolution of a large universe that undergoes a double bounce. The parameters of the model were chosen such that at the bounce point $w \approx -1 \,, \Omega_k=-1\,, a=1$ with a much narrower potential ($\alpha \approx 10^{-5}$). On the right we show the cycle-averaged behaviour of the equation of state parameter, $\overline{w}$, with averaging done across short cycles during fast oscillations, and over short time periods outside of this. In particular, we note that between bounces, there is a period of slow roll from the particle perspective, with $w \approx -1$, however due to the effects of curvature the Hubble parameter evolves rather quickly.}
\label{DB}
\end{center}
\end{figure}

In Fig.~\ref{DB}, 
we see a more complex evolution in which the universe expands, recollapses and then undergoes multiple bounce-recollapse cycles before entering into a de Sitter expansion phase. We can also see the evolution of the scale factor on long time scales where the inflationary expansion is evident. The de Sitter phase in this evolution continues for a large number of $e$-folds, beyond the range of our numerical calculation. We also see a close-up of the behaviour of the Hubble parameter about the primary bounce. Here, we see that the system actually undergoes a very short recollapsing phase before bouncing for a second time. Such behaviour is typical for systems in which the potential is shallow, since the inflaton can quickly exit the inflationary regime, allowing the curvature to overcome the scalar field and causing a recollapse. During this phase, the negative Hubble parameter acts as anti-friction, which accelerates the scalar field allowing it to climb higher up the opposite side of the potential. As long as this effect does not endow the scalar field with more kinetic energy than the height of the potential, the field approaches the plateau in the right phase for a second bounce to occur, followed by a period of inflation. After the bounce, the Hubble parameter is positive, and thus acts as friction on the system which allows for a long period of slow-roll inflation. 

A double-bounce typically occurs as follows: the universe first bounces as the inflaton moving down the potential on the shallow slope of the plateau. After the bounce, there is a short period in which the Hubble parameter grows yet the frictional effect is low. If the phase of the system is such that the inflaton then enters the steeper region of the potential, it can gain sufficient momentum to cross the minimum close to the point of recollapse. The anti-friction drives it up the opposite side of the potential, and the universe bounces for a second time. Usually this second bounce will occur with the inflaton rolling up the potential, and hence it will be followed by a long period of slow roll expansion.  

The double-bouncing scenario (and multiple bounces in general) are of particular importance when considering the amplification of perturbative modes during evolution. This is a phenomenologically interesting feature which runs parallel to that discussed in \cite{Linde:2003hc} in which it was shown that a closed inflationary universe can suppress low multipoles. As we shall see in the following section, we expect to see the longest wavelength modes amplified the most through a bounce. However, the typical behvaiour of a double-bouncing solution is that one bounce is particularly short lived when compared to the frequency of oscillations of the longest wavelengths. Therefore,the phase of recollapse and the second bounce will be effectively smoothed over the longest wavelength modes. However, for a faster oscillating mode, the relative phase of the mode will be important; the significant friction encountered during recollapse whilst in a kinetic phase will lead to suppression of the amplitude, whereas friction during a potential dominated phase will not.  

Before concluding this section, we examine  the possibility that, as the field rises onto the plateau but before turnaround, its variation becomes dominated by its quantum fluctuations. If this were to happen, the field would become oblivious to the scalar potential and the inflaton condensate would spread stochastically, leading to eternal inflation. This is because there would be regions where the quantum ``kicks" would conspire to keep the field on the plateau forever. If this were the case, a double bounce would not be possible. 

Quantum fluctuations can dominate the motion when the classical velocity of the inflaton in field space $|\dot\phi|$ is smaller that the typical quantum fluctuation  
$\delta\phi=H/2\pi$ per Hubble time $\delta t=1/H$. Thus, the criterion in order for the quantum fluctuations to be subdominant is
\be
|\dot\phi|>\frac{\delta\phi}{\delta t}=\frac{H^2}{2\pi}
\;\Rightarrow\;|V'|>\frac{3}{2\pi}H^3,
\label{eternal}
\ee
where we used the slow-roll equation \mbox{$3H\dot\phi=-V'$},
with the prime denoting derivative with respect to the field. Thus, if the slope of the potential is too small, quantum fluctuations take over and eternal inflation occurs. One might think that the above criterion is certainly violated in our case, because at the turnaround point the inflaton stops momentarily and $\dot\phi = 0$. However, it was shown in Ref.~\cite{Dimopoulos:2018upl} that, if the slope is substantial, the duration of the period when $|\dot\phi|<\delta\phi/\delta t$ is less than a Hubble time, which means turnaround occurs before a single quantum ``kick''. In this case, eternal inflation is avoided. The condition for this to happen is none other than Eq.~\eqref{eternal}, provided slow-roll is assumed before turnaround.

How far the inflaton can advance along the potential plateau depends on its initial kinetic density and on the particular scalar potential envisaged. Assuming slow-roll determines the kinetic density in terms of the potential, because of the inflationary attractor. To investigate if the danger of eternal inflation can be avoided in a double bounce scenario, we consider the T-model potential
\be
V(\phi)=\Lambda\tanh^2(\phi/\sqrt{6\alpha}).
\ee
Using that $H=\sqrt{V/3}$ it is straightforward to find that Eq.~\eqref{eternal} is equivalent to
\be
\sinh^2(\phi/\sqrt{6\alpha})<
\frac{4\pi}{\sqrt{2\alpha\Lambda}}.
\ee
The above condition can reduce to
\be
\phi/\sqrt{6\alpha}<\left\{\begin{array}{ll}
\frac12\ln\left(\frac{16\pi}{\sqrt{2\alpha\Lambda}}\right)
& {\rm when}\quad\phi>\sqrt{6\alpha},\\
& \\
\frac{2\sqrt\pi}{(2\alpha\Lambda)^{1/4}}
& {\rm when}\quad\phi<\sqrt{6\alpha}.
\end{array}\right.
\label{criterion}
\ee
The inflationary energy scale is roughly $\Lambda^{1/4}\sim 10^{-2}$. This means that the above criterion for $\phi<\sqrt{6\alpha}$ is always satisfied, when $\alpha<100$ or so. This is expected, for such inflaton values are not on the plateau. In the opposite case, when $\phi>\sqrt{6\alpha}$ (on the plateau now), the criterion in Eq.~\eqref{criterion} becomes $\phi<\sqrt{6\alpha}(6.4-\frac14\ln\alpha)$. Choosing $\alpha=0.01$ we find the bound $\phi<1.8$. Thus, turnaround has to occur at a smaller value of $\phi$ if eternal inflation is to be avoided.

\begin{figure}[htbp]
\subfloat[]{\includegraphics[width=0.48\linewidth]{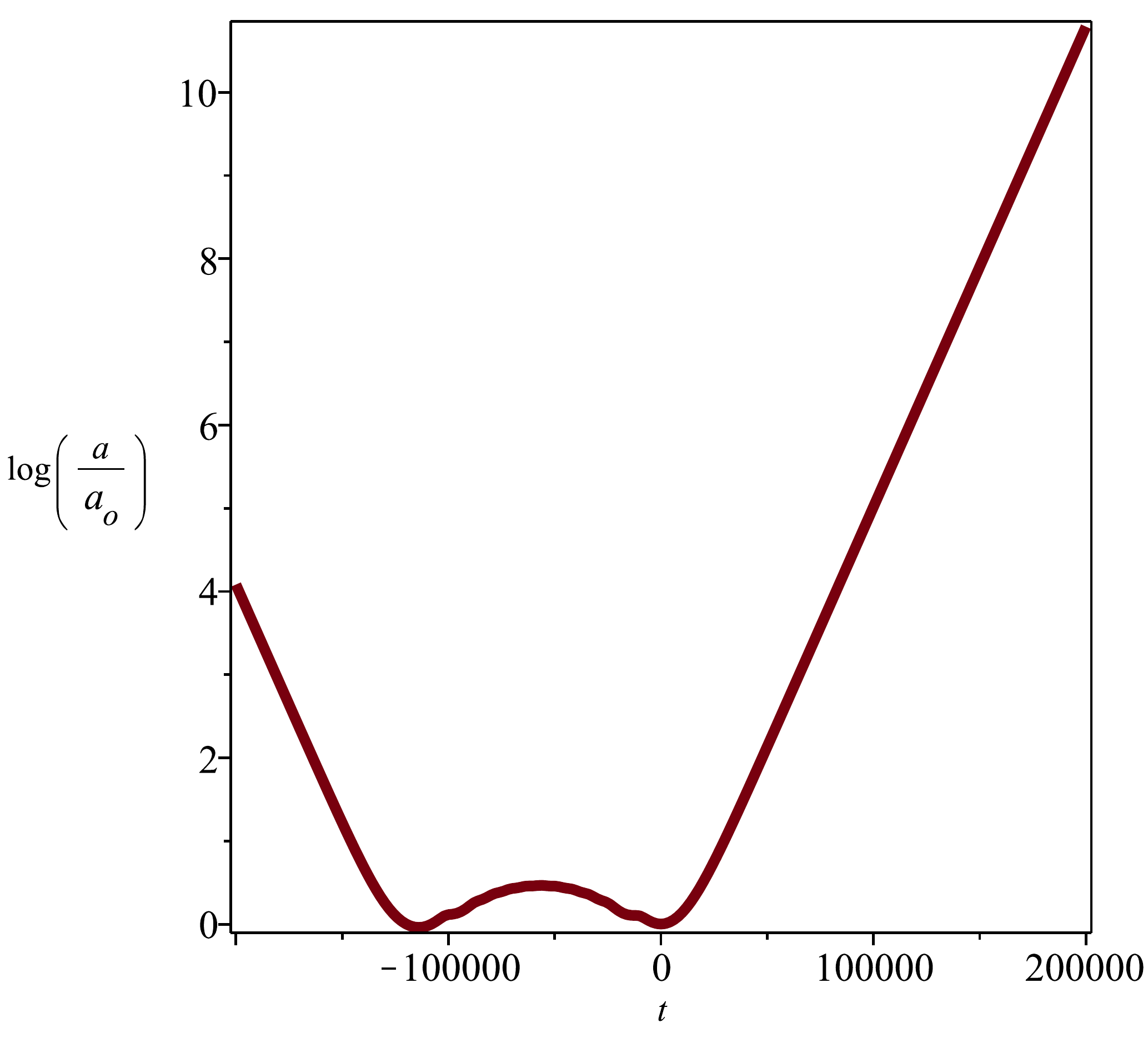}}
\subfloat[]{\includegraphics[width=0.48\linewidth]{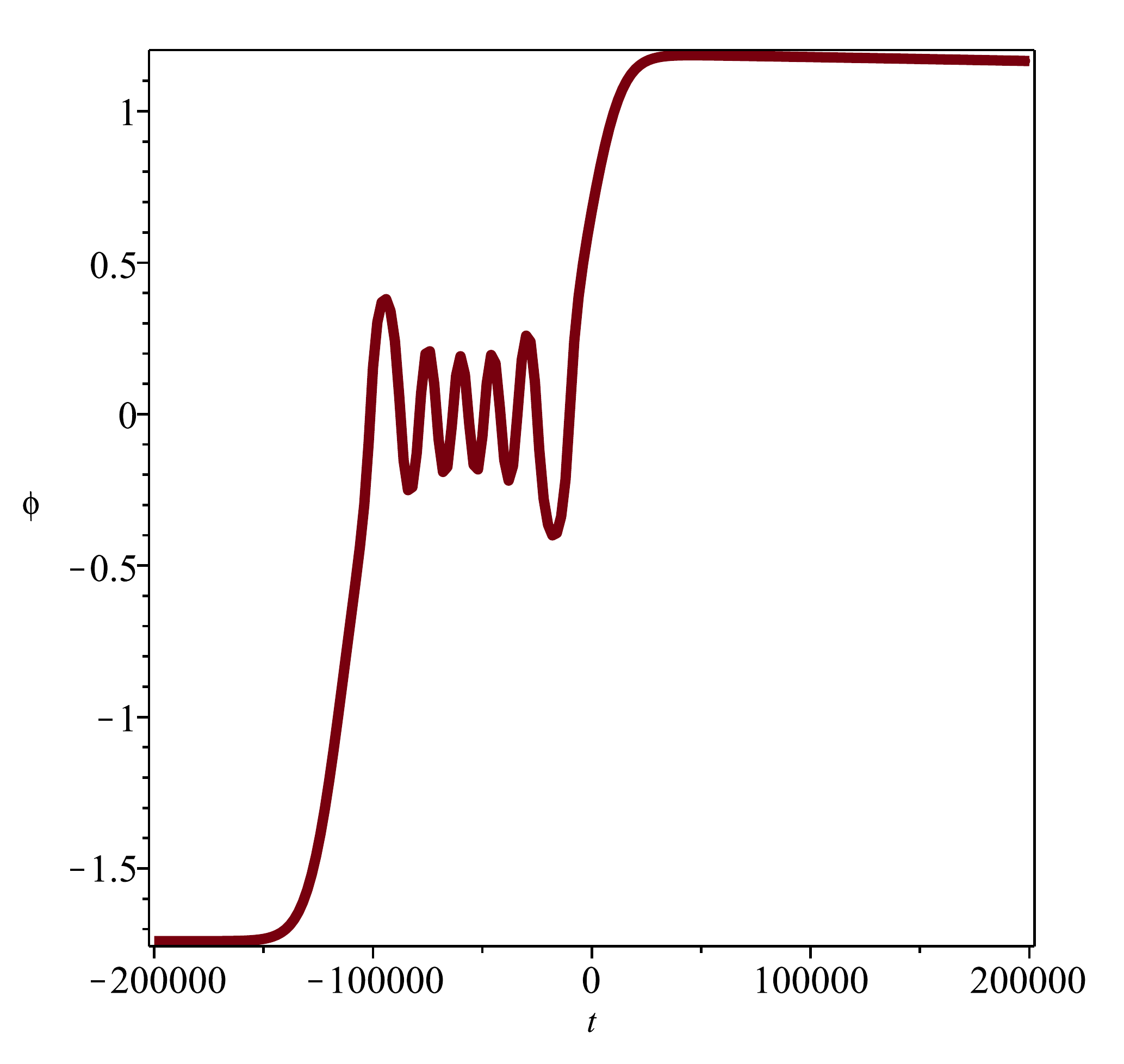}}
\caption{Log of scale factor (a) and scalar field (b) against time for a solution with for the double-bouncing scenario described above. A large universe reduces to a minimal size before undergoing a pair of bounces separated by a short-lived expanding then recollapsing phase. We see that the scalar field avoids the eternal inflation region during the expanding phase that follows the second bounce. This solution continues to inflate for around 90 e-folds.}
\label{RealisticDoubleBounce}
\end{figure}

Through numerical investigation, using the parameters $\Lambda=10^{-8}$, $\alpha=0.01$, and $\Omega_k=-0.005$, we find initial conditions at one bounce that has a double bounce with a large universe at either end. The maximum value of $\phi$ is 1.2, and so is below the eternal inflation value. The system collapses, bounces, recollapses, and then has about 90 e-folds of inflation. So, it is indeed possible to get the double bounce without hitting the eternal inflation problem. The scalar field begins on the plateau of the potential during the collapsing phase and reaches a turning point at the first bounce. This is followed by an oscillating phase during the short-lived expansion and recollapse, and then returning to the plateau for the inflationary phase following the secound bounce. This evolution is shown in Fig.~\ref{RealisticDoubleBounce}.

\begin{figure}[htbp]
\subfloat[]{\includegraphics[width=0.48\linewidth]{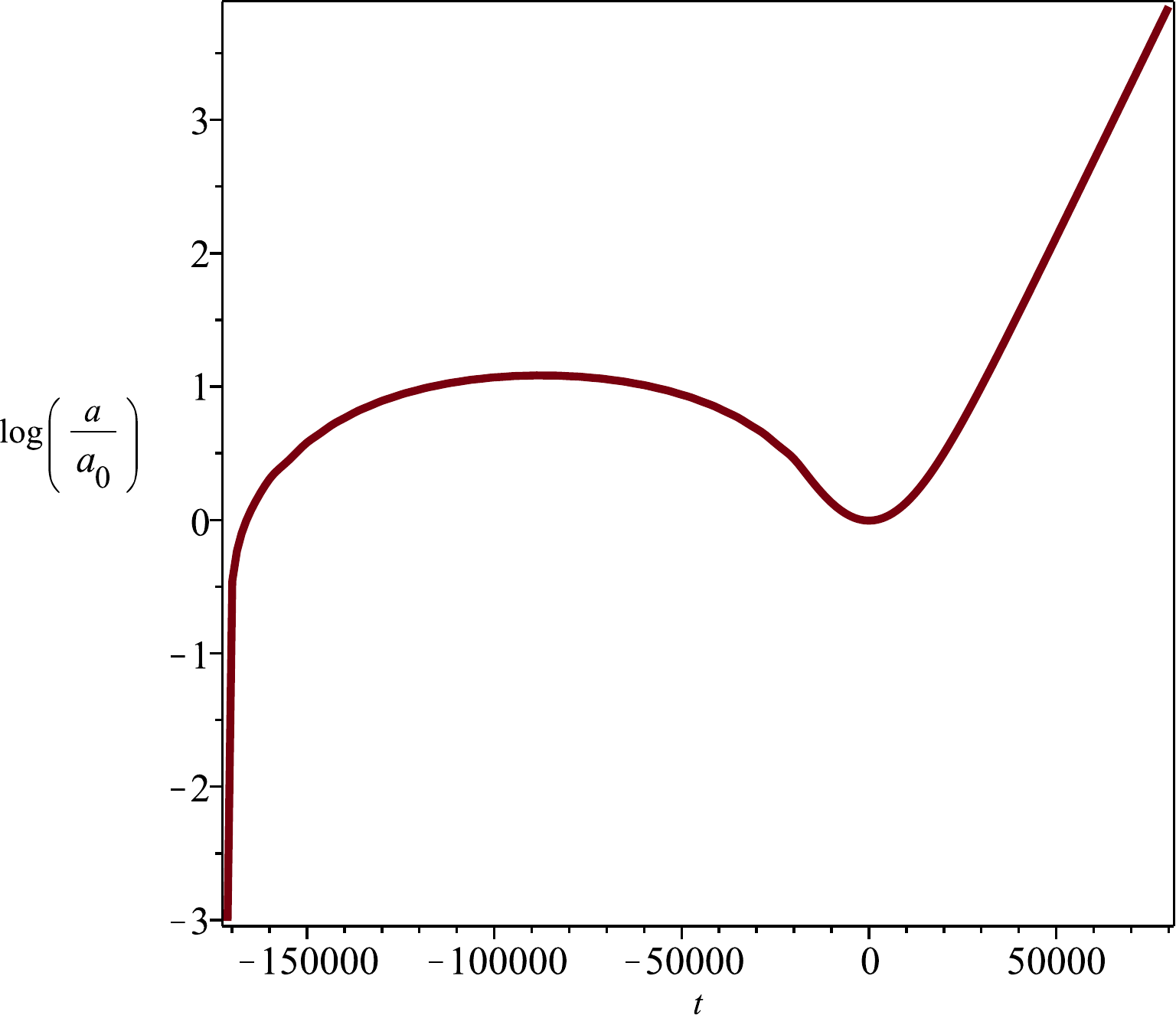}}
\subfloat[]{\includegraphics[width=0.415\linewidth]{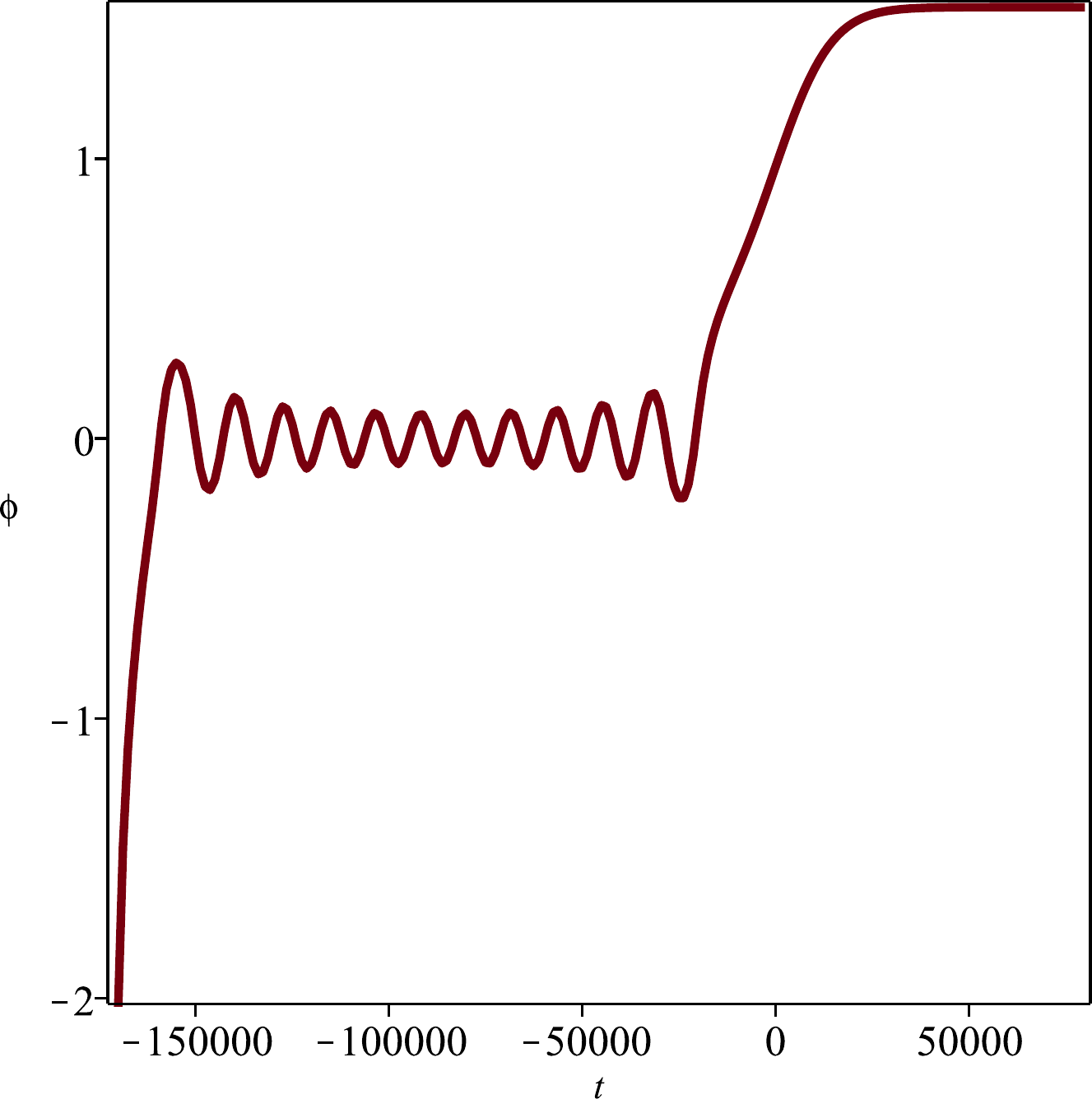}}
\caption{Log of scale factor (a) and scalar field (b) against time for an evolution in which the universe begins at a singularity in the past. The universe in this case expands from an initial singularity to a local maximum size then collapses back to a bounce. Following this bounce the universe then undergoes an extended period of slow-roll inflation (not shown in figure due to focus on singularity and bounce).}
\label{BounceFromNothing}
\end{figure}

Another interesting possibility to consider is that our universe may begin in the far past at a singularity. From this the universe would expand then recollapse to a bounce before undergoing inflation. This was recently discussed in \cite{Matsui:2019ygj} and we direct the reader to the analysis therein for a discussion of the motivation and interesting features of such a model. Such an evolution is indeed possible within the model which we consider and an example numerical evolution is shown in Fig.~\ref{BounceFromNothing}. In such an evolution, the inflaton may be found on the plateau at early times and its kinetic energy tends to infinity as $t \rightarrow -\infty$. However during the initial expansion the inflaton falls down the potential and oscillates about the minimum. During this oscillatory phase the effective averaged equation of state has $\overline{w}=0$, and thus the universe becomes curvature dominated and recollapses. Following this recollapse from a finite maximal scale factor the inflaton can find itself in the right circumstances (low kinetic energy, high potential) for the universe to bounce which can be followed by a prolonged period of inflation. Thus (as discussed in \cite{Matsui:2019ygj}) a universe can `bounce from nothing'. 

\section{Perturbations}
\label{perturbation}
The background model we have thus far described is qualitatively symmetric; for each solution to the equations of motion, there is a counterpart with corresponding dynamics under reversal of time about the bounce point. However, when we consider the evolution of the inhomogeneous modes that evolve on this background, we should not expect to find the same symmetry.  This has been well explored in the case of matter bounce scenarios, an alternative theory to inflation which posits the exit of super-horizon modes during a contracting phase. In particular, non-symmetric phenomena include the generation of black holes, which will persist across any bounces, and the entropic amplification of super-horizon scales as the universe collapses. 

Let us consider the space of scalar perturbations on this background geometry. Recall that we are dealing with a closed FLRW universe, and hence our spatial manifold is $S^3$. We must therefore consider perturbations that consist of hyperspherical harmonics. Following \cite{Bonga:2016iuf,Bonga:2016cje}, we write the eigenfunctions of the Laplace operator on $S^3$ as

\be \nabla^2 Q_{lmn} =  \frac{n^2-1}{r_0^2} Q_{lmn}\;, \ee
where $r_0$ is the comoving radius of curvature.

These constitute the $S^3$ equivalents of Fourier modes. As we take the limit in which the volume of the 3-sphere becomes large compared to the wavelength of mode, these modes are well approximated by the usual Euclidean Fourier modes with wavenumber $k^2=(n^2-1)/r_0^2$.
However, we are particularly interested in the effects of a bounce where the physical volume of the 3-sphere is at its smallest. Furthermore, at the bounce point, the Hubble parameter vanishes exactly. Therefore, the usual frictional term will be absent. This is true regardless of the mode under consideration. An extensive study of the behaviour of perturbations in bouncing cosmologies was performed in Refs.~\cite{Agullo:2012fc,Agullo:2012sh}. Although the analysis was performed in the context of Loop Quantum Cosmology, the results are fairly generic. We will deviate from this type of analysis in two ways: we consider closed cosmologies, and furthermore, we do not set the initial conditions for the semi-classical perturbations at the bounce point. In a quantum context, setting the initial conditions at the bounce point may be a logical choice, since that it is the point of highest energy density, and therefore the point at which quantum effects should be strongest both in terms of matter and geometry. However, we will adopt a different approach here which is better suited to addressing issues of the classical bounce. Since significant portions of phase space do not bounce even in the unperturbed theory, we consider the effects of perturbations to a bouncing background that begin on a collapsing branch rather than setting conditions at the bounce itself. To do otherwise would be to make an undue assumption, namely that the system always bounces. 

Let us first consider scalar curvature perturbations on our background space-time. We take equation (2.17) from Ref.~\cite{Bonga:2016iuf} as the starting point for our analysis, where scalar perturbations are expressed in terms of the variable $q=(\dot{\phi}/H)\zeta$, with $\zeta$ denoting the usual comoving curvature perturbation. Although we will follow this analysis closely, we must deviate in our choice of variables, as~$q$ is ill-defined at the bounce. We therefore use $x=\dot{\phi} \zeta =H q$ as our basic variable, which remains well-defined throughout the evolution of the perturbation. Away from the bounce point, $q$ is easily recovered and we can continue the analysis in the standard inflationary case. Thus, we find our equation of motion for $x$ is given by
\be a_x \ddot{x} + b_x \dot{x} + c_x x =0. \ee
To simplify our analysis, we will write the coefficients $a_x$,$b_x$ and $c_x$  in terms of the Hubble parameter to make it clear which terms become important near the bounce and which terms dominate during slow-roll expansion:
\begin{align}
 \label{scalpert} 
\begin{aligned}
 a_x &=  2H^2(n^2-4) +3 \dot{\phi}^2,   \\
      b_x &=  10H^3(n^2-4)+H\big[  (4n^2+11)\dot{\phi}^2-4(n^2-4)V+6V'\dot{\phi} \big],    \\
      c_x &=  -8H^4(n^2-7)(n^2-4) -2 (\dot{\phi}^2+2V)[(n^2-1)\dot{\phi}^2-4(n^2-4)V]  +24 H(n^2+2)V'\dot{\phi}    \\
             &+ 24V'^2 +12V''\dot{\phi}^2 + 4H^2 \big[(n^4+8n^2+9)\dot{\phi}^2 +2(n^2-4)(n^2-9)V+V'' \big].
\end{aligned}
\end{align}
Although these coefficients are complicated, it is immediately apparent that our system has a well-defined limit when $H$ vanishes. We are therefore able to numerically solve the equations of motion for $x$ through a bounce. 
 
The evolution of tensor perturbations is simpler \cite{Bonga:2016cje}. The dynamical evolution of the amplitude of the transverse, traceless tensor perturbation $h$ of wavenumber $n$ is given by
\be \ddot{h} + 3 H \dot{h} +\left( V+\frac12\dot{\phi}^2-H^2 \right)(n^2-1) h = 0 \label{tenspert} \ee
In both cases, we expect to see only negligible deviations from the usual power spectra generated during the inflationary phase post-bounce, and as such, we expect that even in the presence of a bounce, T-models maintain their attractor nature. We will quantify this statement more carefully in the following section, allowing us to make contact with observations for this class of models. For now, we will primarily concern ourselves with the behaviour of perturbations before the bounce occurs and the backreaction they can have upon our space-time. 

To illustrate the effects of the bounce upon perturbation modes, let us first consider the case in which the bounce occurs when the inflaton can be found on the plateau of the potential. In this case, we neglect the effect of the kinetic energy upon the evolution of the field, and we arrive at the closed de Sitter model whose dynamics are described by
\begin{align}
\label{closeddS}
\begin{aligned}
a &=  a_b \cosh(\lambda t ) \, , 
\\
H &= \lambda\tanh(\lambda t) \,,
\end{aligned}
\end{align}
where $\lambda^2 = V/3$ as set by the value of the potential on the plateau. To simplify the discussion and showcase the qualitative evolution of the different modes, we have set $\Lambda=1=-\Omega_k$. With these assumptions, the evolution of both the tensor and scalar modes becomes identical. The solution for the perturbation variable $x$ can be found analytically as follows, where $P^{\frac{3}{2}}_{n-1/2}$ is the Legendre polynomial of the first kind and  $Q^{\frac{3}{2}}_{n-1/2}$ is the associated Legendre function of the second kind:
\be h = (-\text{sech}^2 \lambda t) ^{3/4} \left[c_1 P_{n-\frac{1}{2}}^{\frac{3}{2}}(\tanh \lambda t )+c_2 Q_{n-\frac{1}{2}}^{\frac{3}{2}}(\tanh \lambda t )\right]. \label{hpertsol} \ee
This solution can be split into two parts, which can broadly be thought of as the components whose amplitude stays fixed and that which changes as the scale factor evolves. As the universe collapses, the amplitude of the evolving mode increases, and conversely, when the universe expands, its amplitude reduces. In a perfectly symmetric bounce such as the one described above, there is no mixing between these modes and thus the amplitudes of oscillations asymptotes towards symmetry. However, when we introduce an asymmetric bounce, the modes mix; a part of the increasing amplitude of the growing mode is transferred into the constant mode (and vice-versa). Therefore, across an asymmetric bounce, we can expect to see that the fluctuations are amplified. Usually, modes outside of the Hubble radius $|H|^{-1}$ freeze out. However, in a contracting universe, Eq.~\eqref{hpertsol} informs us that the amplitude of such modes increases instead. Once the mode is shorter than the Hubble radius, the oscillations dominate over this expansion and so the amplification slows significantly. Thus, we expect that all modes are amplified, and the modes with lowest wavenumber (i.e. the longest wavelength) are amplified the most  \cite{Quintin:2014oea,Quintin:2016qro}.

 \begin{figure}[htbp]
\begin{center}
\subfloat[]{\includegraphics[width=0.48\linewidth]{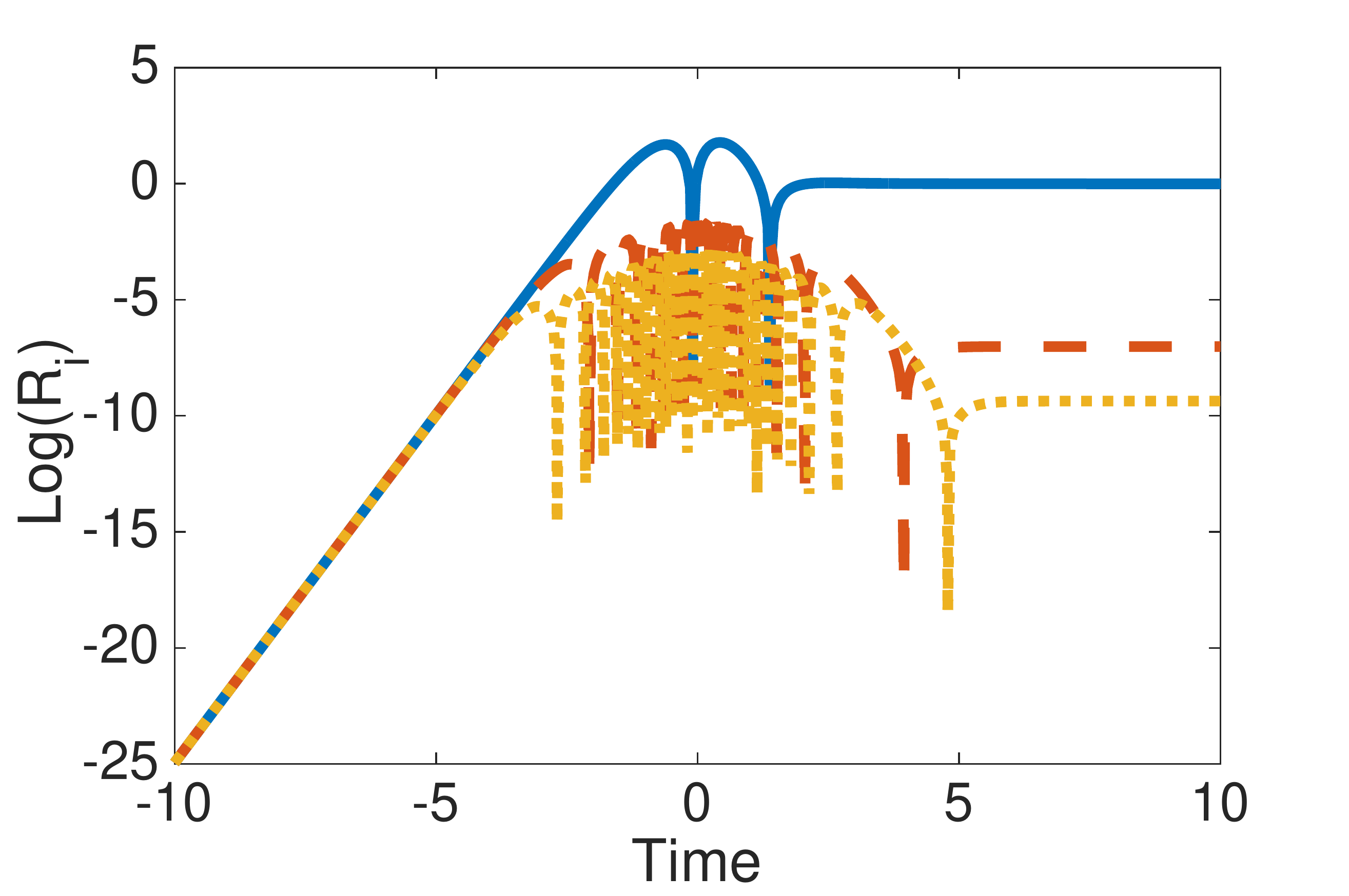}}
\subfloat[]{\includegraphics[width=0.48\linewidth]{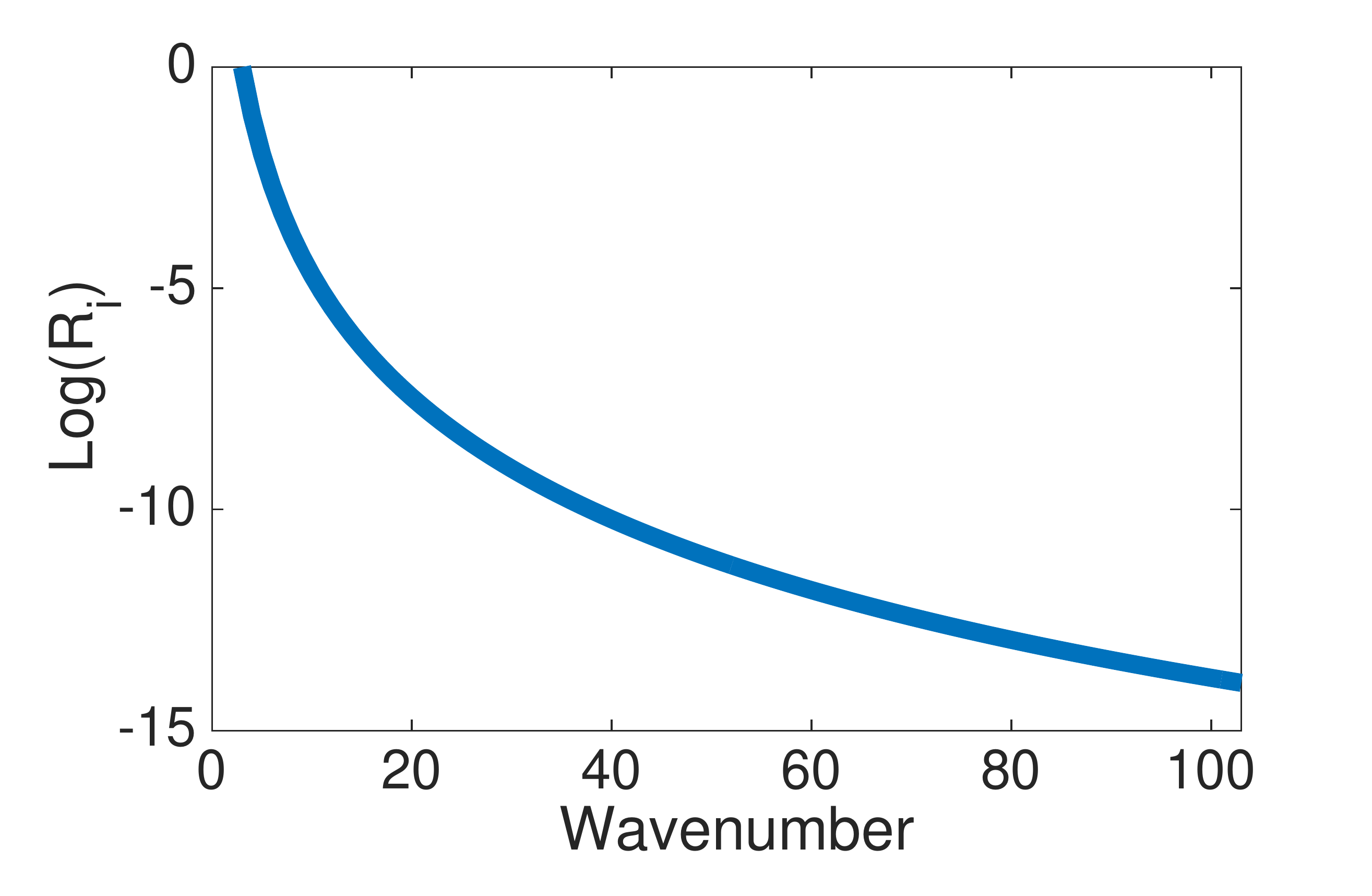}}
 \caption{On the left, the logs of the ratios of amplitudes against time, normalized such that the lowest mode $n=1$ has unit amplitude upon exiting the horizon. Plotted are the behaviours of the $n=3,18$ and $33$ (solid blue, dashed red and dotted yellow respectively). On the right, the logs of ratios of the amplitude increases of the wavemodes of tensor and scalar perturbations across a bounce to that of the lowest mode ($n=3$). We see that the largest increases are found at the low end of the spectrum, with the distribution becoming smaller and flatter as the wavenumber increases. We should expect to see a red tilt to the spectrum of perturbations, but the amplitude of this will strongly depend on the total number of $e$-folds following the bounce; large expansions will suppress this effect significantly.}
\label{ratio}
\end{center}
\end{figure}

\section{Contact with observations}
\label{observ}

We now turn to examine how the phenomenology of a double-bounce scenario fares in the context of inflation. Recall that if the background dynamics allow for a double bounce, we expect that soon after the second bounce, the Hubble parameter will reach an approximately constant value, and the universe will be described well by a de Sitter approximation. As noted in the previous section, perturbations will be enhanced as the universe evolves through the bounce. However, as we will show in this section, we expect that the power spectrum for smaller modes (which inflate for a larger number of $e$-folds) will not will not deviate from the usual results post-bounce. As a result, we will demonstrate that it is possible to realize a double-bounce model with predictions in good agreement with current observations. For the rest of this section, we will restore units of $M_P^{-1}=\sqrt{8\pi G}$.

We first note that post-bounce, de Sitter space is achieved very quickly as long as the field starts reasonably high on the plateau, as we can see in Fig.~\ref{DB}. Since the number of $e$-folds is defined as $N = - H \, dt$ and $H = 0$ at the bounce, we expect the number of $e$-folds until (quasi-)de Sitter is achieved to be almost negligible. The deviation from de Sitter is encoded in $\epsilon$, with $\epsilon = 0$ corresponding to exact de Sitter. Using the solutions through the bounce \eqref{closeddS}, we find that $|\epsilon|$ becomes smaller than $|\epsilon|_{\rm dS}$ at a number of $e$-folds given by
\begin{align}
N_{\rm dS} = \ln \cosh \csch^{-1} |\epsilon|_{\rm dS}.
\end{align}
We can set $|\epsilon|_{\rm dS}= 1$ as a hard limit for quasi-de Sitter. This returns $N_{\rm dS} = 0.367$, whereas $|\epsilon|_{\rm dS}= 0.1$ returns $N_{\rm dS} = 2.31 $. Therefore, generically after the second bounce, the universe will rapidly approach the flat de Sitter evolution within ${\rm O}(1)$ $e$-folds: only scales that leave the horizon before that are going to see anything other than a quasi-de Sitter universe.

For large scales, we expect two distinct behaviours: first, we expect finite-size effects to become important as the scales approach the radius of the unverse and second, we expect the effects of the bounce to cause a deviation from the usual near scale invariance, since large scales leave the horizon before de Sitter is achieved. Finite size effects are known to become important at large scales, affecting the power spectrum. 
However, the presence of the bounce further modifies the usual treatment, regardless of the spatial curvature of the universe. As noted above, the gauge-independent perturbation $q = \delta\phi + (\dot\phi/H) \zeta$ is ill-defined in a bounce. However, we may instead work in the uniform curvature gauge, where $\zeta = 0$. In this case, we consider the variable $v= a \delta \phi$, which remains well-defined through the bounce and follows the equation of motion~\cite{Allen:2004vz}
   \begin{align}
  \label{pertclosedeq1}
v_n'' + \left(  \frac{n^2 -1}{r_0^2} - \frac{a''}{a} \right) u_n =0.
\end{align}
where the prime indicates differentiation with respect to the conformal time $\eta = \int dt/a$. 
This is nothing more than collection of a simple harmonic oscillators with a time-dependent masses, which means that the definition of a vacuum is time-dependent as well. In the de Sitter case, $a''/a \approx 6/\eta^2$ close to the singularity, which means that a Euclidean vacuum (the well-known Bunch--Davies initial condition) can be imposed. However, close to the bounce where $a \propto \cosh \lambda t $ and $t =  2 \lambda^{-1} \tanh ^{-1}\left[\tan(\lambda\eta/2)\right]$, we find that in the early limit:
   \begin{align}
  \label{pertclosedeq2}
v_n'' + \left(  \frac{n^2 -1 }{r_0^2}  - \frac{1}{ \lambda^{-2} + \frac{\eta^2 }{6}}  \right) u_n =0.
\end{align}
Therefore, we can apply the Bunch--Davies vacuum condition at the bounce if all modes become \emph{nearly} massless at the bounce, which is the case for small $\lambda$ (we show this below). The physical meaning of this condition can be thought of as follows: the bounce ``smoothes out'' the perturbations as the horizon becomes infinitely large, since there is no typical length scale. This is akin to assuming a (nearly) maximally symmetric space for observationally relevant scales.

In the domain where $\eta \lambda \gg 1 $ (which occurs soon after the bounce but before smaller modes leave the horizon), the solution to this equation is the usual superposition of the Bessel functions 
 \begin{align}
v_n = c_1 \sqrt{n\eta } \, J_{\frac{\sqrt{3}}{2}}\left[\frac{(n^2-1) \eta }{r_0}\right]+c_2 \sqrt{n \eta } Y_{\frac{\sqrt{3}}{2}}\, \left[\frac{(n^2-1) \eta }{r_0}\right].
\end{align}
We impose the usual normalisation condition $u'u^*-{u^*}'u = i$ and the requirement that at some early time just after the bounce the oscillations were free $v \propto e^{\pm ik\eta}$. Then, identifing $\delta\phi$ as the comoving curvature perturbation $\mathcal{R}$ in the zero-curvature gauge, and since $\braket{\delta\phi|\delta\phi} \sim H^2$ along with $\mathcal{R} = H\, \delta\phi/\dot \phi$, we arrive at the well-known expression for the dimensionless spectrum of scalar perturbations:
  \begin{align} 
  \mathcal{P}_\zeta = \frac{1}{8\pi^2 M_P^2}\frac{H^4}{\dot\phi^2} \approx \frac{1}{8\pi^2 M_P^2} \frac{H^2}{\epsilon},
\end{align}
where the normalisation comes from the usual CMB transfer functions.

As noted above, we expect a deviation from the nearly flat spectrum for scales that exit the horizon roughly early enough after the bounce, before quasi-de Sitter is reached but after the initial conditions can be assumed to be Euclidean. First, we note that the COBE normalisation may be used to estimate the value of $\lambda M_P =\sqrt{V_0/3}$ of the potential. Using
\be
\frac{H^2}{M_P^2 \epsilon} = 1.59 \times 10^{-6},
\ee
and $\epsilon \sim 10^{-2}$ at the pivot scale, we may estimate $\lambda = 4.21\times 10^{-5} M_P $.  Note that for this value of~$\lambda$, the scales in which there is departure from de Sitter but the initial conditions are valid are set by the scale $ a_b/\lambda \approx 10^{30} \ {\rm Mpc}$ for $a_b \sim e^{-N_*}$, where the $\Lambda$CMB evolution of the scale factor is dwarfed by the early exponential expansion (compare with the radius of the observable universe  $14.3 \ {\rm Gpc}$). Therefore, we are certainly safe in assuming time-independent vacua as initial conditions.  

Using the closed de Sitter expression for the scale factor, the dimensionless spectrum looks like
  \begin{align} 
\mathcal{P}_\zeta= - \frac{\lambda^2}{M_P^2}  \sinh^2 (\lambda t)  \tanh^2 ( \lambda t). 
\end{align}
Horizon crossing occurs at $k = a H$ as usual, which gives 
  \begin{align} 
k = a_b \lambda  \sinh (\lambda  t).
\end{align}
Therefore, we find
  \begin{align} 
\mathcal{P}_\zeta =   \frac{k^2}{M_P^2 a_b^2} \tanh^2  \left[{\rm arcsinh} \left( \frac{k}{a_b \lambda}\right) \right]    =  \frac{\lambda^2}{M_P^2}-\frac{k^2} {M_P^2 a_b^2} + \mathcal{O} (k^4/a_b^4), 
\end{align}
where we expand for large wavelengths for which the above assumption of time-independent vacua holds.

If we assume that outside of largest observable scales, the de Sitter approximation quickly broke down (i.e. assuming the bare minimum e-folds of growth), then the scales where we expect to see deviation from near scale invariance correspond to $e^{-N_{\rm dS}} r_{\rm CMB}$, where $r_{\rm CMB} = 14.3 \ {\rm Gpc}$. This in turn corresponds to an angular scale of $\theta = e^{-N_{\rm dS}}/(2\pi )$, varying from $6^\circ$ to $0.9^\circ$ for $N_{\rm dS} \in [0.367, 2.31]$ as estimated above. This corresponds to multipoles of  $\ell \lesssim 180$ to  $\ell \lesssim 30$; in particular, a suppression to the latter, corresponding to a slightly slower relaxation to de Sitter, agrees with observations. Note that the relative correction to the spectrum corresponding to the lowest multipoles is of the order
  \begin{align} 
\frac{\Delta \mathcal{P}_\zeta }{\mathcal{P}_\zeta}   = - \frac{k^2}{a_b^2 \lambda^2} = - \frac{ \ell^2/r_{\rm CMB}^2}{e^{-2 N_{\rm tot}} \lambda^2}.
\end{align}
By varying the total number of $e$-foldings $N_{\rm tot}$, it is possible to arrive at a percentage-level modification of the order of $0.1$ to $1 \%$ to the spectrum. Such an effect is largely model independent, insofar as it only requires a bounce followed by a relaxation to de Sitter, and may explain current observations. In fact,  matching the suppression with observations can lead to limits on the number of $e$-folds: for instance, in the above example, $N_{\rm tot}\gtrsim 125$ is specifically ruled out, as it predicts a much stronger suppression than observed at low multipoles.

Note that this effect is complementary to the finite-curvature effects that can cause an enhancement of the power spectra at low $n$, dominating over the enhancement effect.  Scales with $n \gg 3$ quickly find themselves deep in the Hubble horizon. The pivot scales $k = 0.05 \ {\rm Mpc}^{-1}$ and $k = 0.002 \ {\rm Mpc}^{-1}$ correspond to $n=3152$ and $n=126$ respectively for the smallest Universe compatible with observations \cite{Bonga:2016cje}. Therefore we expect that the dependence of the vacuum on the wavenumber is negligible, and the power spectra are only modified in an observationally relevant way due to the bounce, not finite-curvature effects. Moreover, the suppression is not contrary to the enhancement of perturbations through the bounce found in the previous section, since we have set the initial conditions close the bounce and not at horizon exit.
It is true that perturbations are amplified through a bounce, but as the comoving horizon becomes infinite and all modes can talk to each other, we may reasonably assume that gauge-invariant perturbations approach a maximally symmetric set of initial conditions. 
%In general, the suppression of effects is only observable for the lowest of modes$ $

Having discussed the implications of the bounce on the low multipole modes, we now turn our attention to the predictions of our $\alpha$-attractor potential for scales which experience de Sitter for their entire lifetime. We remind that our model is governed by the Lagrangian
  \begin{align}
\sqrt{-g} \, \mathcal{L} = \frac12M_P^2 R - \frac{1}{2} (\partial\phi)^2 - 3\alpha\, m^2 M_P^2  \tanh^2\frac{\phi}{\sqrt{6\alpha}M_P},
\end{align}
where we have reinstated $\alpha$ in the potential for dimensional consistency. We recall the usual expressions for the tilt of the spectrum and the tensor-to-scalar ratio:
\begin{align}
  \begin{aligned} 
n_s  &= 1 -2\epsilon + \eta,
\\
r &= 16 \epsilon,
\end{aligned}
\end{align}
where
\begin{align}
  \begin{aligned} 
\epsilon &= \frac12M_P^2 \left(\frac{V'}{V}\right)^2,
\\
\eta &= M_P^2 \frac{\epsilon'}{\epsilon} \frac{V'}{V},
\end{aligned}
\end{align}
where the prime now denotes derivative with respect to the inflaton field. These definitions for the slow-roll parameters agree with the Hubble slow-roll parameters $\epsilon_H = -\dot H/H^2$ and $\eta_H = \dot \epsilon_H/(H\epsilon_H)$ in the limit of slow-roll.

As noted in the previous section, the plateau shape of the potential, which is generic to attractor models, will give rise to results that should match observations quite well. Calculating the slow-roll parameters, we find
  \begin{align}
\epsilon &= \frac{4  }{3 \alpha }    \csch ^2   \left ( \sqrt{\frac{2}{3\alpha} }\phi/M_P  \right),
\\
\eta &= \frac{2 }{3 \alpha }\left[1 + \coth ^2\left(\frac{\phi/M_P }{\sqrt{6\alpha}  }\right) \right]  \text{sech}^2 \left(\frac{\phi/M_P }{\sqrt{6\alpha}} \right).
\end{align}
The value of $\phi$ at the end of inflation is found as usual by setting $\epsilon =1$, returning
  \begin{align}
\phi_{\rm end}=  \sqrt{\frac{3 \alpha}{2}} M_P\sinh ^{-1}\left(\frac{2}{\sqrt{3\alpha}  }\right) .
\end{align}
With this result, the number of $e$-folds is related to the field value $\phi$ as
  \begin{align}
\phi = \sqrt{\frac{3\alpha}{2}} M_P  \cosh ^{-1}\left(
   \sqrt{1+\frac{4}{3\alpha } }    
+\frac{ 4  N }{3 \alpha }
\right) .
\end{align}
We may therefore easily calculate the value of the spectral index and the tensor-to-scalar ratio:
  \begin{align}
n_s  &= \frac{1}{6 \alpha } \left[\left(\sqrt{\frac{12}{\alpha }+9}-3\right) \alpha +4 N-8\right] \ \text{csch}^2\left[\frac{1}{2} \cosh ^{-1}\left(\sqrt{\frac{4}{3 \alpha }+1}+\frac{4 N}{3 \alpha }\right)\right],
\\
r &= \frac{48 \alpha }{4 N^2+2 \alpha  N\sqrt{\frac{12}{\alpha }+9} \, + 3 \alpha }.
\end{align}
In particular, the current constraint on the tensor-to-scalar ratio $r<0.06$ sets the allowed range of $\alpha$ as follows:
  \begin{align}
0.06 \ge \frac{48 \alpha }{4 N^2+2 \alpha  N\sqrt{\frac{12}{\alpha }+9} \, + 3 \alpha },
\end{align}
returning $\alpha < 33.5$. Moreover, for the scalar spectrum, we find
  \begin{align}
\mathcal{P}_\zeta = \frac{3 \alpha^2  m^2}{4\pi^2M_P^2} \sinh ^4\left[\frac{1}{2} \cosh ^{-1}\left(\sqrt{\frac{4}{3 \alpha }+1}+\frac{4 N}{3 \alpha }\right)\right].
\end{align}
The value of $N$ depends on the details of reheating, but we expect it to be very close to $N=60$ for the pivot scale. Using the COBE normalisation $\mathcal{P}_\zeta = (2.099\pm 0.029)\times 10^{-9}$, we may find a relation between $\alpha$ and $m$. For small values of $\alpha$, we find that the value of $m$ is mostly insensitive to the model, as we would expect from an attractor: we find $m \approx 7.25 \times 10^{-6}\,M_P$. As a result, in order to avoid the curvature effects from being dominated by the potential, we set $m^2 \lesssim 6 \alpha\,M_P^2$, which sets $\alpha> 8.78 \times 10^{-12}$. This gives us the window for $\alpha$ as
  \begin{align}
10^{-11}\lesssim \alpha \lesssim 34.
\end{align}
As a result, the predicted values for $n_S$ and $r$ are
  \begin{align}
  \begin{aligned}
0.9667  &\lesssim  n_s   \lesssim 0.9669,
\\
10^{-14} &< r   \lesssim 0.06.
\end{aligned}
\end{align}
For large values of $\alpha$, roughly $10 \lesssim \alpha \lesssim 33$, the predictions are at slight tension with observations, as they depend on the particulars of the model. However, for small $\alpha$, the attractor nature of the theory becomes manifest and the predictions of the model are well within the Planck contour.

While such a model is concordant with observations following a bounce, we must finally ask whether a model with these parameters is consistent with the observed curvature in the universe today. As noted in the previous section, at the bounce we have $a_b =M_P^2 \sqrt{-\Omega_k/\Lambda}$. This can be used to find $\Omega_k$ in the convention that $a = 1$ at the bounce, which roughly corresponds to a difference in convention of roughly $e^{-2N_{\rm tot}}$ (thanks to the small evolution of the curvature after inflation subsides). However, if $a=1$ is taken to hold today and $a = a_b$ at the bounce, we may find that the value of the curvature today is
 \begin{align}
\Omega_k = \frac{\Lambda e^{-2N_{\rm tot}} }{M_P^4} ,
 \end{align}
 where we used the relation $\Lambda  = 3 \alpha m^2 M_P^2$.
 For the range $1.58 \times 10^{-21}\lesssim \Lambda/M_P^4 \lesssim 1.57 \times 10^{-9} $ (which is realised for the allowed values of $\alpha$), we find that the spatial curvature today indeed set to be very small but nonzero nonetheless for these models. This tells us that a bouncing cosmology followed by a period of inflation is particularly robust: it can be realised without requiring extremely large (negative) curvature close to the bouncing point, and once a double bounce is realised,  the curvature will be driven down to small levels as usual,  in agreement with observations.

\section{Discussion}
\label{conclusions}

We have proposed a bouncing scenario in the presence of a plateau potential stemming from a conformal action featuring a ${\rm SO}(1,1)$ symmetry, allowing a single conformally couple field to provide both the inflationary dynamics and the dominant matter content at the bounce. We demonstrated that in a contracting universe, a bounce can be realised quite generically in the $\alpha$-attractor paradigm when potential domination is satisfied and the field is sufficiently high up on the plateau. Performing a phase space analysis, we have shown that the measure on the space of solutions is dominated by bouncing solutions when equipped with an appropriate cut-off due to large field values being physically indistinguishable. By performing both analytic and numeric evolutions, we demonstrate that across the bounce, low wavenumber modes of perturbations are amplified. However, if the primordial perturbations at observable are normalised with respect to a time-independent maximally symmetric vacuum at early times, we find that there will be a percent-level suppression at low multipoles, which agrees with CMB observations. This is a distinct feature from the suppression that occurs due to finite size effects. At large multipoles, the de Sitter evolution is reached, and we find that the predictions of the double bounce model agree with those of standard attractor inflation for the Planck pivot scale.

We should briefly comment on an implicit assumption which we made to simplify
our treatment. We have considered a roughly homogeneous Universe before the
bounce, despite the fact that a contracting universe would amplify
inhomogeneities, which could threaten the bounce. However, this danger of
excessive inhomogeneities can be alleviated in a number of ways. One possibility
is considering an early stage of inflation before the bounce, which can take the
system from the Planck scale down to the T-model plateau. This can be
introduced by another field, such that the plateau lies actually at the bottom
of the potential valley, whose walls rise up to the Planck density. Examples of
this kind of structure in the context of $\alpha$-attractors can be seen in
Refs.~\cite{Carrasco:2015rva,Dimopoulos:2016yep}, where the early inflation is
power-law and lasts for only a short duration. Similarly, but without the need for an additional
field, a slight modification of the inflaton potential can achieve a limited
early inflation stage, by lifting the valley towards the Planck scale (see for
example Fig.~5 of Ref.~\cite{Linde:2017pwt}).%
\footnote{In terms of $\alpha$-attractors, we may consider a change in the
  kinetic term from \mbox{$1/[1- \phi^2/(6\alpha)]$} to
  \mbox{$1/[1- \phi^2/(6\alpha)\exp(-\lambda\phi)]$}, essentially ``smearing''
  out'' the poles.
%This deforms the manifold to such an extent that it is no longer Kähler: its
%hyperbolic geometry is only apparent near the origin.
  Solving the equations of motion,
%in closed form becomes incredibly difficult in this case, but
  it is possible to show that due to the canonicalisation, any kinetic term
  which is roughly this shape, produces an inflection-point type plateau
  irrespective of the potential. By tuning $\alpha$ and $\lambda$, it
  is possible to arrive at a steep "drop", corresponding to a long but
  finite plateau, which then rises up and makes contact with the Planck scale.}
One other way, which would not invoke early inflation, is to consider that the
universe topology is not trivial, such that homogenisation is achieved through
``chaotic mixing'' \cite{Ellis:1970ey,Cornish:1996st,Linde:2004nz} due to the
fact that the entire compact universe comes in causal contact soon after its
quantum nucleation. Needless to say, slow-roll inflation after the bounce
may easily enlarge the boundaries of the compact universe to superhorizon scales
today, so that any trace of non-trivial topology is lost. Finally, one could
also perceive the bounce acting as a filter on excessive inhomogeneity. If we
start with a lot of universes contracting, then those that are not homogeneous
enough would not experience a bounce. Therefore, only the roughly homogeneous
ones would end make it through to inflation and expand like our Universe. How likely or unlikely the
above scenarios are, touches upon the infamous measure problem of inflationary
cosmology \cite{Corichi:2013kua,Gibbons:2006pa,Sloan:2015bha,Sloan:2016nnx}, which goes beyond the scope of our~paper.

A possible extension of our analysis could include multiple fields, 
%or similarly, a field on a higher dimensional potential. 
acting on separate scales for the bounce and the subsequent de Sitter expansion of the universe. A simple example would be the potential $V=\Lambda_1 \tanh^{2n} (\frac{\theta}{\sqrt{6\alpha}})+\Lambda_2 \tanh^{2m} (\frac{\phi}{\sqrt{6\beta}})$, where the two scales could allow separate scales at which the bounce and inflation could occur. This effect would be particularly pronounced if $m \gg n$ such the potential is very steep and narrow in the direction of highest energy.

Note that if the bounce is marginal (i.e. we bounce at $w=-1/3$ and nearby $w>-1/3$), we expect there to be singularities just beyond the horizon. Specifically, if we find ourselves in a patch which has $w<-1/3$ while $w>-1/3$ outside it, everything beyond the horizon becomes singular. If we find ourselves in the only patch that has this feature, the universe will have closed up on a point.

\section*{Acknowledgements}

KD is supported (in part) by the Lancaster-Manchester-Sheffield Consortium for Fundamental Physics under STFC grant: ST/L000520/1.
DS is grateful for useful discussion with Beatrice Bonga and Brajesh Gupt relating to the evolution of perturbations. The authors are grateful to the anonymous referee from whose insight we benefited in revising this manuscript.

\appendix
\section{The hyperbolic tangent oscillator}
\label{hyptan}
\begin{figure}[htbp]
\begin{center}
\subfloat[]{\includegraphics[width=0.48\linewidth]{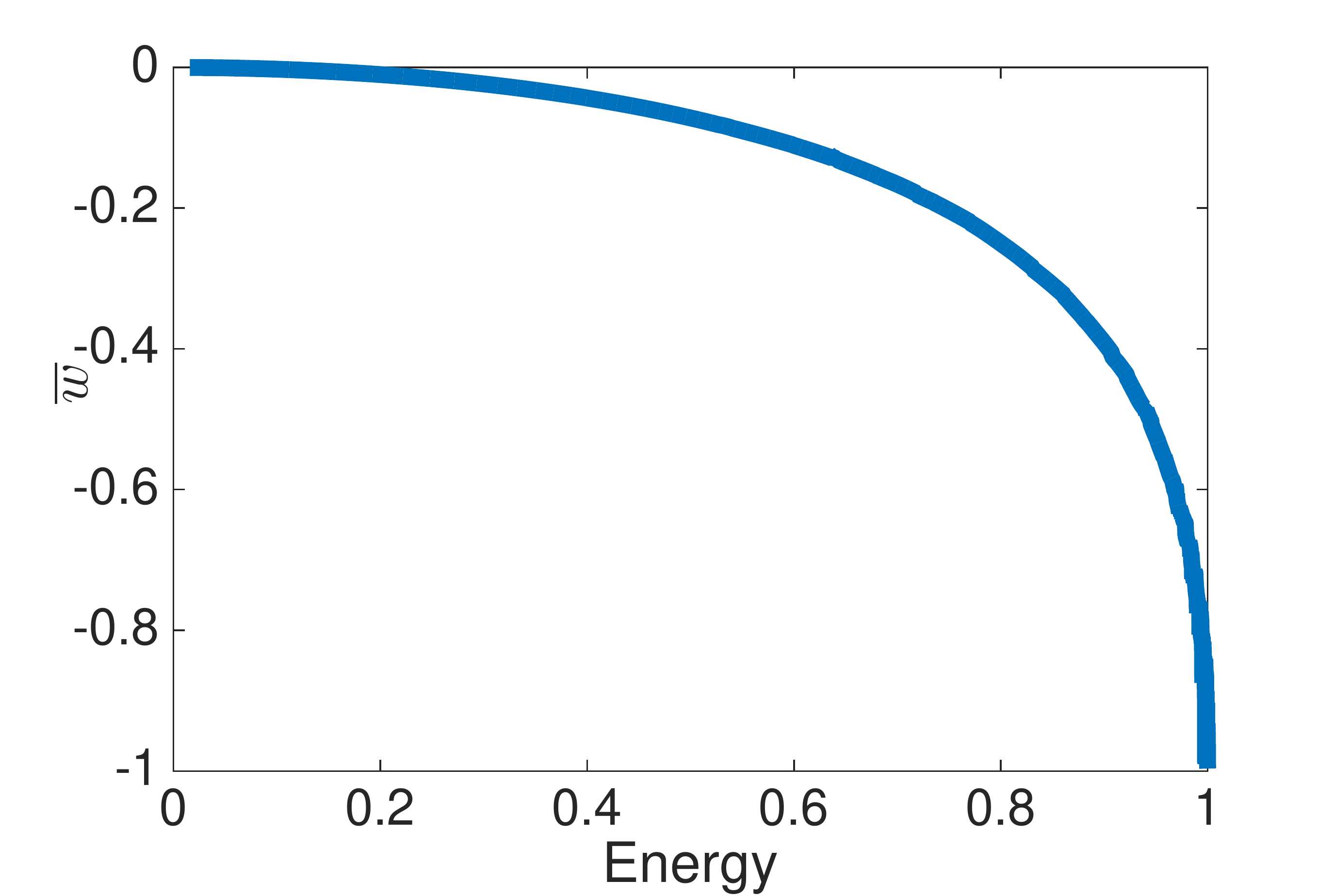}}
\subfloat[]{\includegraphics[width=0.48\linewidth]{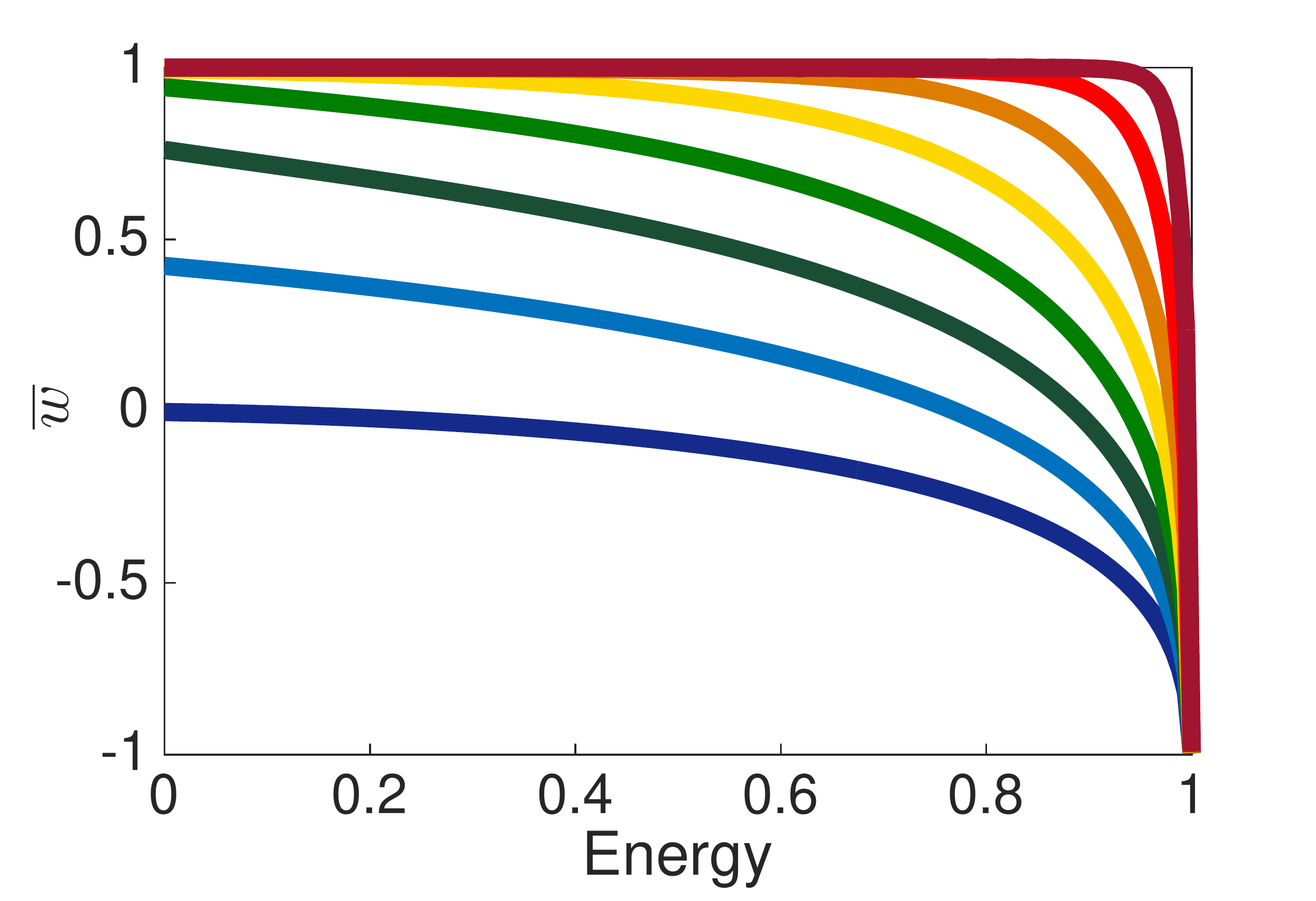}}
 \caption{The average value of $w$ across a cycle against the energy of the oscillator for increasing values of~$n$ in the potential. Shown on the left~(a) is the behaviour of~$w$ in the case of a quadratic $\tanh$ potential. Shown on the right~(b) is the same for a range of even monomials. To show the overall behaviour, the lines are each $2^m$ for $m$ between 1 and 7, with the colours chosen such that the blue line corresponds to $m=2$ and~$m$ increases as we move towards the red end of the spectrum.  Here we see that as $n$ increases, the average value of $w$ - $\overline{w}$ increases, asymptoting to $\overline{w}=1$ (corresponding to the square well). When $n=2$, the results coincide with the above behaviour, and the maximum value of $\overline{w}$ is 0. For higher $n$, this value increases, as does the range of energies over which $\overline{w}$ is close to 1. }
\label{wavg}
\end{center}
\end{figure}

Throughout this paper, we have considered a scalar field whose potential was $\tanh^2(x)/2$ (here we include the factor of $1/2$ to facilitate an easier comparison with the usual $x^2/2$ potential of a harmonic oscillator). Let us now establish some of the relevant properties of such a field. The Hamiltonian for the field in the absence of gravitational coupling (and hence Hubble friction) is simply
\be \mathcal{H} = \frac12P_\phi^2 + \frac12\tanh^2(\phi) = E . \ee
Here we have fixed the size of the potential to unity without loss of generality. The equation of motion for our system is
\be  \label{STO} 
\ddot{\phi} = - \tanh (\phi) \sech^2 (\phi). \ee
The first thing to notice is that the qualitative behaviour of the system is now dependent upon the total energy $E$. In the case where $E \gg 1$, the system appears as that of a free field as the potential has little effect on the motion. When $E\ll 1$, the system behaves like a harmonic oscillator. For a harmonic oscillator, the kinetic energy and the potential follow $\cos^2(t)$  and $\sin^2(t)$ respectively. Hence, 
\be w_{SHO} = \frac{KE-PE}{KE+PE} = \frac{\cos^2(t) - \sin^2(t)}{\cos^2(t)+\sin^2(t)} = \cos(2t) .\ee 
As a result, across a complete cycle, the average of $w$ is zero. However, when we consider the behaviour of the $\tanh$ potential, we come across a subtlety. When $E>1$, the field no longer oscillates since the system is unbound. Hence, the equivalent behaviour is captured by the long-term average of $w$, which tends to $w_u = \frac{E-2}{E}$. When $E<1$, the average is difficult to find analytically. In the limit where $E \ll 1$, the harmonic oscillator becomes a good approximation to the dynamics of the system and hence the average across a cycle is zero. For $0<E<1$, we can find the average behaviour from numerical averaging shown in Fig/~\ref{wavg}.

We can observe from Fig.~\ref{wavg} that when $E>\sqrt{3}/2$ we find that the average of $w<-1/3$ However, we note that this approximation would only hold in a limit in which we could neglect the Hubble friction, which would not be valid in this regime; in other words $\ddot{a}<0$ whenever the field is oscillating with sufficiently quickly that we can use the average field approximation. Hence a bounce that occurs during the oscillating phase of the field will not yield a large universe; a recollapse will occur before an oscillation is complete. We are therefore motivated to consider bounces that occur when the field is strongly damped. 

Let us now turn to the case where we consider higher powers of $\tanh(\phi)$ to form our potential. In this situation, we note that taking the limit as $n\rightarrow\infty$ yields no potential at all, since the $\tanh(\phi)<1$ for all finite values of $\phi$. However, by changing the width of the potential we can produce a more reasonable function to use; $V_n = \tanh^n (k \phi)$ for even integers $n$. In particular, consider fixing the potential at a given point irrespective of the value of $n$ - as an example let us fix the value of $V_n$ at $\phi=\phi_o$:
\be V_n = \frac12\tanh^n (k \phi) = \frac12\tanh^2(\phi_o). 
\ee
From this prescription we find that $k$ is given:
\be k = \frac{\tanh ^{-1}\left[\tanh ^{2/n}(\phi_o)\right]}{\phi_o}. \ee
It is apparent that in the infinite limit, $\phi_o$ will be the width of our square well, as the potential can only pass through the fixed value at the wall.  

The equation of motion for our inflaton is now
\be \ddot{\phi} = -\frac{k n}{2} \sech^2 (k \phi) \tanh^{n-1} (k \phi). \ee
Fig.~\ref{wavg} was generated by numerically solving this system for a range of energies of the oscillator. We see that as the power $n$ increases, the behaviour of the system approaches more and more closely that of a square well. In particular, we note that the average behaviour of the field across an oscillation is not well approximated by dust, but becomes better and better approximated by that of a stiff fluid ($p=\rho$).

\end{document}